\documentclass[12pt]{article}

\begin{document}

\author{Gunn Alex Quznetsov\thanks{%
quznets@yahoo.com}}
\title{A Piece of the Lepton Theory from a Probability}
\maketitle

\begin{abstract}
A masses of a leptons deduced from a representation of a probability density
vector by a spinors. A massive W and Z bosons and a massless A boson are
obtained from a transformations for which a density vector is invariant.
\end{abstract}

\allowbreak I use the following denotations:

\[
1_2=\left[ 
\begin{array}{cc}
1 & 0 \\ 
0 & 1
\end{array}
\right] \mbox{, }0_2=\left[ 
\begin{array}{cc}
0 & 0 \\ 
0 & 0
\end{array}
\right] 
\]

and for $k\geq 2$:

\[
1_{2k}=\left[ 
\begin{array}{cc}
1_k & 0_k \\ 
0_k & 1_k
\end{array}
\right] \mbox{, }0_{2k}=\left[ 
\begin{array}{cc}
0_k & 0_k \\ 
0_k & 0_k
\end{array}
\right] \mbox{.} 
\]

\[
\sigma _1=\left[ 
\begin{array}{cc}
0 & 1 \\ 
1 & 0
\end{array}
\right] \mbox{, }\sigma _2=\left[ 
\begin{array}{cc}
0 & -i \\ 
i & 0
\end{array}
\right] \mbox{, }\sigma _3=\left[ 
\begin{array}{cc}
1 & 0 \\ 
0 & -1
\end{array}
\right] 
\]

are the Pauli matrices.

The Clifford pentad \cite{MD} $\stackrel{\circ }{\beta }$ is:

\begin{equation}
\begin{array}{c}
\beta _1=\left[ 
\begin{array}{cc}
\sigma _1 & 0_2 \\ 
0_2 & -\sigma _1
\end{array}
\right] ,\beta _2=\left[ 
\begin{array}{cc}
\sigma _2 & 0_2 \\ 
0_2 & -\sigma _2
\end{array}
\right] ,\beta _3=\left[ 
\begin{array}{cc}
\sigma _3 & 0_2 \\ 
0_2 & -\sigma _3
\end{array}
\right] , \\ 
\gamma _0=\left[ 
\begin{array}{cc}
0_2 & 1_2 \\ 
1_2 & 0_2
\end{array}
\right] =\beta _5,\beta _4=i\left[ 
\begin{array}{cc}
0_2 & 1_2 \\ 
-1_2 & 0_2
\end{array}
\right] \mbox{, }
\end{array}
\label{a0}
\end{equation}

\[
\beta _0=\left[ 
\begin{array}{cc}
1_2 & 0_2 \\ 
0_2 & 1_2
\end{array}
\right] \mbox{, }\gamma _5=\left[ 
\begin{array}{cc}
1_2 & 0_2 \\ 
0_2 & -1_2
\end{array}
\right] \mbox{.} 
\]

\section{Masses}

Let

\[
\left\langle \rho \left( t,x,y,z\right) ,j_x\left( t,x,y,z\right) ,j_y\left(
t,x,y,z\right) ,j_z\left( t,x,y,z\right) \right\rangle 
\]

be a probability current 3+1 vector field \cite{PRB} and $\psi \left(
t,x,y,z\right) $ be any complex spinor field:

\[
\psi =\left| \psi \right| \left[ 
\begin{array}{c}
\exp \left( i\stackrel{*}{\gamma }\right) \cos \left( \stackrel{*}{\beta }%
\right) \cos \left( \stackrel{*}{\alpha }\right) \\ 
\exp \left( i\stackrel{*}{\theta }\right) \sin \left( \stackrel{*}{\beta }%
\right) \cos \left( \stackrel{*}{\alpha }\right) \\ 
\exp \left( i\stackrel{*}{\varphi }\right) \cos \left( \stackrel{*}{\chi }%
\right) \sin \left( \stackrel{*}{\alpha }\right) \\ 
\exp \left( i\stackrel{*}{\upsilon }\right) \sin \left( \stackrel{*}{\chi }%
\right) \sin \left( \stackrel{*}{\alpha }\right)
\end{array}
\right] \mbox{.} 
\]

In this case the following system of equations:

\begin{equation}
\left\{ 
\begin{array}{c}
\psi ^{\dagger }\psi =\rho \mbox{,} \\ 
\psi ^{\dagger }\beta _1\psi =j_x\mbox{,} \\ 
\psi ^{\dagger }\beta _2\psi =j_y\mbox{,} \\ 
\psi ^{\dagger }\beta _3\psi =j_z
\end{array}
\right|  \label{a1}
\end{equation}

has got the following form:

\[
\left\{ 
\begin{array}{c}
\psi ^{\dagger }\psi =\rho \mbox{,} \\ 
\left| \psi \right| ^2\left( 
\begin{array}{c}
\cos ^2\left( \stackrel{*}{\alpha }\right) \sin \left( 2\stackrel{*}{\beta }%
\right) \cos \left( \stackrel{*}{\theta }-\stackrel{*}{\gamma }\right) - \\ 
-\sin ^2\left( \stackrel{*}{\alpha }\right) \sin \left( 2\stackrel{*}{\chi }%
\right) \cos \left( \stackrel{*}{\upsilon }-\stackrel{*}{\varphi }\right)
\end{array}
\right) =j_x\mbox{,} \\ 
\left| \psi \right| ^2\left( 
\begin{array}{c}
\cos ^2\left( \stackrel{*}{\alpha }\right) \sin \left( 2\stackrel{*}{\beta }%
\right) \sin \left( \stackrel{*}{\theta }-\stackrel{*}{\gamma }\right) - \\ 
-\sin ^2\left( \stackrel{*}{\alpha }\right) \sin \left( 2\stackrel{*}{\chi }%
\right) \sin \left( \stackrel{*}{\upsilon }-\stackrel{*}{\varphi }\right)
\end{array}
\right) =j_y\mbox{,} \\ 
\left| \psi \right| ^2\left( \cos ^2\left( \stackrel{*}{\alpha }\right) \cos
\left( 2\stackrel{*}{\beta }\right) -\sin ^2\left( \stackrel{*}{\alpha }%
\right) \cos \left( 2\stackrel{*}{\chi }\right) \right) =j_z\mbox{.}
\end{array}
\right| 
\]

Hence for every probability current vector $\left\langle \rho
,j_x,j_y,j_z\right\rangle $: the spinor $\psi $, obeyed to this system,
exists.

The operator $\widehat{U}\left( t,\triangle t\right) $, which acts in the
set of these spinors, is denoted as the evolution operator for the spinor $%
\psi \left( t,x,y,z\right) $, if:

\[
\psi \left( t+\triangle t,x,y,z\right) =\widehat{U}\left( t,\triangle
t\right) \psi \left( t,x,y,z\right) \mbox{.} 
\]

$\widehat{U}\left( t,\triangle t\right) $ is a linear operator.

The set of the spinors, for which $\widehat{U}\left( t,\triangle t\right) $
is the evolution operator, is denoted as the operator $\widehat{U}\left(
t,\triangle t\right) $ space.

The operator space is the linear space.

Let for an infinitesimal $\triangle t$:

\[
\widehat{U}\left( t,\triangle t\right) =1+i\triangle t\widehat{H}\left(
t\right) \mbox{.} 
\]

Hence for an elements of the operator $\widehat{U}\left( t,\triangle
t\right) $ space:

\[
i\widehat{H}=\partial _t\mbox{.} 
\]

If the functions $\rho $, $j_x$, $j_y$, $j_z$ fulfill to the continuity
equation \cite{PRB}:

\[
\partial _t\rho +\partial _xj_x+\partial _yj_y+\partial _zj_z=0 
\]

then:

\[
\left( \left( \partial _t\psi ^{\dagger }\right) \beta _0+\left( \partial
_x\psi ^{\dagger }\right) \beta _1+\left( \partial _y\psi ^{\dagger }\right)
\beta _2+\left( \partial _z\psi ^{\dagger }\right) \beta _3\right) \psi = 
\]

\[
=-\psi ^{\dagger }\left( \left( \beta _0\partial _t+\beta _1\partial
_x+\beta _2\partial _y+\beta _3\partial _z\right) \psi \right) \mbox{.} 
\]

Let:

\[
\widehat{Q}=\left( i\widehat{H}+\beta _1\partial _x+\beta _2\partial
_y+\beta _3\partial _z\right) \mbox{.} 
\]

Hence:

\[
\psi ^{\dagger }\widehat{Q}^{\dagger }\psi =-\psi ^{\dagger }\widehat{Q}\psi %
\mbox{.} 
\]

Hence $i\widehat{Q}\left( t,x,y,z\right) $ is the Hermitian for the matrix
product operator.

Hence a real functions $\varphi _{i,j}\left( t,x,y,z\right) $ and $\varpi
_{i,j}\left( t,x,y,z\right) $ for which:

\[
-i\widehat{Q}= 
\]

\[
\left[ 
\begin{array}{cccc}
\varphi _{1,1} & \varphi _{1,2}+i\varpi _{1,2} & \varphi _{1,3}+i\varpi
_{1,3} & \varphi _{1,4}+i\varpi _{1,4} \\ 
\varphi _{1,2}-i\varpi _{1,2} & \varphi _{2,2} & \varphi _{2,3}+i\varpi
_{2,3} & \varphi _{2,4}+i\varpi _{2,4} \\ 
\varphi _{1,3}-i\varpi _{1,3} & \varphi _{2,3}-i\varpi _{2,3} & \varphi
_{3,3} & \varphi _{3,4}+i\varpi _{3,4} \\ 
\varphi _{1,4}-i\varpi _{1,4} & \varphi _{2,4}-i\varpi _{2,4} & \varphi
_{3,4}-i\varpi _{3,4} & \varphi _{4,4}
\end{array}
\right] 
\]

exist.

Let $G_t$, $G_z$, $K_t$ and $K_z$ are the solution of the following system
of equations:

\[
\left\{ 
\begin{array}{c}
G_t+G_z+K_t+K_z=\varphi _{1,1}\mbox{,} \\ 
G_t-G_z+K_t-K_z=\varphi _{2,2}\mbox{,} \\ 
G_t-G_z-K_t+K_z=\varphi _{3,3}\mbox{,} \\ 
G_t+G_z-K_t-K_z=\varphi _{4,4};
\end{array}
\right| 
\]

$G_x$ and $K_x$ are the solution of the following system of equations:

\[
\left\{ 
\begin{array}{c}
G_x+K_x=\varphi _{1,2}\mbox{,} \\ 
-G_x+K_x=\varphi _{3,4}\mbox{;}
\end{array}
\right| 
\]

$G_y$ and $K_y$ are the solution of the following system of equations:

\[
\left\{ 
\begin{array}{c}
-G_y-K_y=\varpi _{1,2}\mbox{,} \\ 
G_y-K_x=\varpi _{3,4}\mbox{.}
\end{array}
\right| 
\]

In this case:

\[
\begin{array}{c}
-i\widehat{Q}= \\ 
=\left( G_t\beta _0+G_x\beta _1+G_y\beta _2+G_z\beta _3\right) + \\ 
+\left( K_t\beta _0+K_x\beta _1+K_y\beta _2+K_z\beta _3\right) \gamma _5+
\end{array}
\]

\[
+\left[ 
\begin{array}{cccc}
0 & 0 & \varphi _{1,3}+i\varpi _{1,3} & \varphi _{1,4}+i\varpi _{1,4} \\ 
0 & 0 & \varphi _{2,3}+i\varpi _{2,3} & \varphi _{2,4}+i\varpi _{2,4} \\ 
\varphi _{1,3}-i\varpi _{1,3} & \varphi _{2,3}-i\varpi _{2,3} & 0 & 0 \\ 
\varphi _{1,4}-i\varpi _{1,4} & \varphi _{2,4}-i\varpi _{2,4} & 0 & 0
\end{array}
\right] \mbox{.} 
\]

If

\[
\left\{ 
\begin{array}{c}
\left( M_0+M_{z,0}\right) =\varphi _{1,3}\mbox{,} \\ 
\left( M_0-M_{z,0}\right) =\varphi _{2,4}\mbox{,}
\end{array}
\right| 
\]

\[
\left\{ 
\begin{array}{c}
\left( M_4+M_{z,4}\right) =\varpi _{1,3}\mbox{,} \\ 
\left( M_4-M_{z,4}\right) =\varpi _{2,4}\mbox{,}
\end{array}
\right| 
\]

\[
\left\{ 
\begin{array}{c}
\left( M_{x,0}+M_{y,4}\right) =\varphi _{1,4}\mbox{,} \\ 
\left( M_{x,0}-M_{y,4}\right) =\varphi _{2,3}\mbox{,}
\end{array}
\right| 
\]

\[
\left\{ 
\begin{array}{c}
\left( M_{x,4}+M_{y,0}\right) =\varpi _{1,4}\mbox{,} \\ 
\left( M_{x,4}-M_{y,0}\right) =\varpi _{2,3}
\end{array}
\right| 
\]

then

\[
\left[ 
\begin{array}{cccc}
0 & 0 & \varphi _{1,3}+i\varpi _{1,3} & \varphi _{1,4}+i\varpi _{1,4} \\ 
0 & 0 & \varphi _{2,3}+i\varpi _{2,3} & \varphi _{2,4}+i\varpi _{2,4} \\ 
\varphi _{1,3}-i\varpi _{1,3} & \varphi _{2,3}-i\varpi _{2,3} & 0 & 0 \\ 
\varphi _{1,4}-i\varpi _{1,4} & \varphi _{2,4}-i\varpi _{2,4} & 0 & 0
\end{array}
\right] = 
\]

\[
=M_0\gamma _0+M_4\beta _4-M_{x,0}\gamma _\zeta ^0-M_{x,4}\zeta
^4+M_{y,0}\gamma _\eta ^0+M_{y,4}\eta ^4-M_{z,0}\gamma _\theta
^0-M_{z,4}\theta ^4\mbox{;} 
\]

here $\gamma _\zeta ^0$, $\zeta ^4$, $\gamma _\eta ^0$, $\eta ^4$, $\gamma
_\theta ^0$, $\theta ^4$ are the chromatic pentads \cite{QD}, \cite{HC}
members and $\gamma _0$ and $\beta _4$ is the light pentad $\stackrel{\circ 
}{\beta }$ members. Since in this paper I will not consider a quarks then
everywhere below: 
\[
M_{x,0}=M_{x,4}=M_{y,0}=M_{y,4}=M_{z,0}=M_{z,4}=0
\]

hence:

\[
\begin{array}{c}
-i\widehat{Q}= \\ 
=\left( G_t\beta _0+G_x\beta _1+G_y\beta _2+G_z\beta _3\right) + \\ 
+\left( K_t\beta _0+K_x\beta _1+K_y\beta _2+K_z\beta _3\right) \gamma _5+ \\ 
+M_0\gamma _0+M_4\beta _4\mbox{.}
\end{array}
\]

$\left\{ \beta _1,\beta _2,\beta _3,\beta _4,\gamma _0\right\} $ is the
Clifford pentad.

If $j_x=\rho u_x$, $j_y=\rho u_y$, $j_z=\rho u_z$ then $u_x$, $u_y$, $u_z$
are the components of the average velocity. Hence $\beta _1$, $\beta _2$, $%
\beta _3$ define the components of the average velocity (\ref{a1}).

If

\[
j_{x_5}=\psi ^{\dagger }\gamma _0\psi \mbox{, }j_{x_4}=\psi ^{\dagger }\beta
_4\psi \mbox{, }j_{x_5}=\rho u_{x_5}\mbox{, }j_{x_4}=\rho u_{x_4} 
\]

then

\[
\begin{array}{c}
u_{x_5}=\sin \left( 2\stackrel{*}{\alpha }\right) \left[ 
\begin{array}{c}
\cos \left( \stackrel{*}{\beta }\right) \cos \left( \stackrel{*}{\chi }%
\right) \cos \left( \stackrel{*}{\gamma }-\stackrel{*}{\varphi }\right) + \\ 
+\sin \left( \stackrel{*}{\beta }\right) \sin \left( \stackrel{*}{\chi }%
\right) \cos \left( \stackrel{*}{\theta }-\stackrel{*}{\upsilon }\right)
\end{array}
\right] \mbox{,} \\ 
u_{x_4}=\sin \left( 2\stackrel{*}{\alpha }\right) \left[ 
\begin{array}{c}
\cos \left( \stackrel{*}{\beta }\right) \cos \left( \stackrel{*}{\chi }%
\right) \sin \left( \stackrel{*}{\gamma }-\stackrel{*}{\varphi }\right) + \\ 
+\sin \left( \stackrel{*}{\beta }\right) \sin \left( \stackrel{*}{\chi }%
\right) \sin \left( \stackrel{*}{\theta }-\stackrel{*}{\upsilon }\right)
\end{array}
\right]
\end{array}
\]

and if $\rho \neq 0$ then

\begin{equation}
u_x^2+u_y^2+u_z^2+u_{x_5}^2+u_{x_4}^2=1.  \label{vel}
\end{equation}

From \cite{PG} the maximal velocity of the information propagation in the
space-time is 1.

Hence of only all five elements of the Clifford pentad lends the entire kit
of the velocity components and, for the completeness, yet two ''space''
coordinates $x_5$ and $x_4$ should be added to our three $x,y,z$.

Let

\[
\begin{array}{c}
\mathbf{\Psi }\left( t,x,y,z,x_5,x_4\right) = \\ 
=\psi \left( t,x,y,z\right) \exp \left( -i\left( x_5M_0\left( t,x,y,z\right)
+x_4M_4\left( t,x,y,z\right) \right) \right) \mbox{.}
\end{array}
\]

In this case the motion equation is the following:

\begin{equation}
\begin{array}{c}
\beta _0i\partial _t\mathbf{\Psi }+\beta _1i\partial _x\mathbf{\Psi }+\beta
_2i\partial _y\mathbf{\Psi }+\beta _3i\partial _z\mathbf{\Psi }+\gamma
_0i\partial _{x_5}\mathbf{\Psi }+\beta _4i\partial _{x_4}\mathbf{\Psi }+ \\ 
+\left( G_t\beta _0+G_x\beta _1+G_y\beta _2+G_z\beta _3\right) \mathbf{\Psi }%
+ \\ 
+\left( K_t\beta _0+K_x\beta _1+K_y\beta _2+K_z\beta _3\right) \gamma _5%
\mathbf{\Psi }=0
\end{array}
\label{gkk}
\end{equation}

Let a evolution operator $\widehat{U}\left( t,\triangle t\right) $ be
denoted as {\it a Planck evolution operator} if a tiny positive real
number $h$ and a functions $N_\varphi \left( t,x,y,z\right) $ and $N_\varpi
\left( t,x,y,z\right) $, having a range of values in the set of the integer
numbers, exist for which:

\[
M_0=N_\varphi h\mbox{ and }M_4=N_\varpi h\mbox{.} 
\]

Let $-\frac \pi h\leq x_5\leq \frac \pi h$ , $-\frac \pi h\leq x_4\leq \frac
\pi h$,

$\mathbf{\Psi }\left( t,x,y,z,\pm \frac \pi h,x_4\right) =0$ and $\mathbf{%
\Psi }\left( t,x,y,z,x_5,\pm \frac \pi h\right) =0$.

In this case the Fourier series for $\mathbf{\Psi }$ is of the following
form:

\[
\begin{array}{c}
\mathbf{\Psi }\left( t,x,y,z,x_5,x_4\right) = \\ 
=\psi \left( t,x,y,z\right) \sum_{\nu ,\kappa }\delta _{-\nu ,N_\varphi
\left( t,x,y,z\right) }\delta _{-\kappa ,N_\varpi \left( t,x,y,z\right)
}\exp \left( -ih\left( \nu x_5+\kappa x_4\right) \right) \mbox{.}
\end{array}
\]

Here:

\begin{eqnarray*}
\delta _{-\nu ,N_\varphi } &=&\frac h{2\pi }\int_{-\frac \pi h}^{\frac \pi
h}\exp \left( ih\left( \nu x_5\right) \right) \exp \left( iN_\varphi
hx_5\right) dx_5=\frac{\sin \left( \pi \left( \nu +N_\varphi \right) \right) 
}{\pi \left( \nu +N_\varphi \right) }\mbox{,} \\
\delta _{-\kappa ,N_\varpi } &=&\frac h{2\pi }\int_{-\frac \pi h}^{\frac \pi
h}\exp \left( ih\left( \kappa x_4\right) \right) \exp \left( iN_\varpi
hx_4\right) dx_4=\frac{\sin \left( \pi \left( \kappa +N_\varpi \right)
\right) }{\pi \left( \kappa +N_\varpi \right) }\mbox{.}
\end{eqnarray*}

If denote:

\[
\phi \left( t,x,y,z,-\nu ,-\kappa \right) =\psi \left( t,x,y,z\right) \delta
_{\nu ,N_\varphi \left( t,x,y,z\right) }\delta _{\kappa ,N_\varpi \left(
t,x,y,z\right) } 
\]

then

\[
\begin{array}{c}
\mathbf{\Psi }\left( t,x,y,z,x_5,x_4\right) = \\ 
=\sum_{\nu ,\kappa }\phi \left( t,x,y,z,\nu ,\kappa \right) \exp \left(
-ih\left( \nu x_5+\kappa x_4\right) \right) \mbox{.}
\end{array}
\]

From the properties of $\delta $ in every point $\left\langle
t,x,y,z\right\rangle $: either

\[
\mathbf{\Psi }\left( t,x,y,z,x_5,x_4\right) =0 
\]

or an integer numbers $\nu _0$ and $\kappa _0$ exist for which:

\begin{equation}
\mathbf{\Psi }\left( t,x,y,z,x_5,x_4\right) =\phi \left( t,x,y,z,\nu
_0,\kappa _0\right) \exp \left( -ih\left( \nu _0x_5+\kappa _0x_4\right)
\right) \mbox{.}  \label{dlt}
\end{equation}

That is for the every space-time point: either this point is empty or single
mass is placed in this point.

Let on the space of these spinors the scalar product $\mathbf{\Phi }*\mathbf{%
\Psi }$ be denoted as the following:

\[
\mathbf{\Phi }*\mathbf{\Psi }=\left( \frac h{2\pi }\right) ^2\int_{-\frac
\pi h}^{\frac \pi h}dx_5\int_{-\frac \pi h}^{\frac \pi h}dx_4\cdot \left( 
\mathbf{\Phi }^{\dagger }\mathbf{\Psi }\right) \mbox{.} 
\]

In this case:

\[
\mathbf{\Psi }*\beta _\mu \mathbf{\Psi }=\psi ^{\dagger }\beta _\mu \psi %
\mbox{.} 
\]

for $0\leq \mu \leq 3$

Hence from (\ref{a1}):

\[
\left\{ 
\begin{array}{c}
\mathbf{\Psi }^{\dagger }*\mathbf{\Psi }=\rho \mbox{,} \\ 
\mathbf{\Psi }^{\dagger }*\beta _1\mathbf{\Psi }=j_x\mbox{,} \\ 
\mathbf{\Psi }^{\dagger }*\beta _2\mathbf{\Psi }=j_y\mbox{,} \\ 
\mathbf{\Psi }^{\dagger }*\beta _3\mathbf{\Psi }=j_z\mbox{.}
\end{array}
\right| 
\]

\subsection{Bi-zero-nonzero-mass state}

Let

\[
\begin{array}{c}
\mathbf{\Psi }\left( t,x,y,z,x_5,x_4\right) = \\ 
=\phi \left( t,x,y,z,0,0\right) +\phi \left( t,x,y,z,n,k\right) \exp \left(
ih\left( nx_5+kx_4\right) \right) \mbox{.}
\end{array}
\]

Let $\epsilon _\mu $ ($1\leq k\leq 4$) be a basis in which pentad $\stackrel{%
\circ }{\beta }$ has got a form (\ref{a0}) and let

\begin{equation}
\begin{array}{c}
\mathbf{\Psi }\left( x_5,x_4\right) = \\ 
=\sum_{r=1}^4\phi _r\left( 0,0\right) \epsilon _r+\exp \left( -ih\left(
nx_5+kx_4\right) \right) \sum_{k=1}^4\phi _k\left( n,k\right) \epsilon _k
\end{array}
\label{bir}
\end{equation}

Hence in the basis

\[
\left\langle \epsilon _r,\exp \left( -ih\left( nX+kY\right) \right) \epsilon
_k\right\rangle : 
\]

a 8-components bi-spinor:

\[
\Psi =\left[ 
\begin{array}{c}
\phi _1\left( 0,0\right) \\ 
\phi _2\left( 0,0\right) \\ 
\phi _3\left( 0,0\right) \\ 
\phi _4\left( 0,0\right) \\ 
\phi _1\left( n,k\right) \\ 
\phi _2\left( n,k\right) \\ 
\phi _3\left( n,k\right) \\ 
\phi _4\left( n,k\right)
\end{array}
\right] 
\]

corresponds to $\mathbf{\Psi }$.

From (\ref{dlt}): in every point $\left\langle t,x,y,z\right\rangle $:

\begin{equation}
\Psi =\left[ 
\begin{array}{c}
\phi _1\left( 0,0\right) \\ 
\phi _2\left( 0,0\right) \\ 
\phi _3\left( 0,0\right) \\ 
\phi _4\left( 0,0\right) \\ 
0 \\ 
0 \\ 
0 \\ 
0
\end{array}
\right] \mbox{ or }\Psi =\left[ 
\begin{array}{c}
0 \\ 
0 \\ 
0 \\ 
0 \\ 
\phi _1\left( n,k\right) \\ 
\phi _2\left( n,k\right) \\ 
\phi _3\left( n,k\right) \\ 
\phi _4\left( n,k\right)
\end{array}
\right]  \label{dl2}
\end{equation}

of $\delta $ characteristics.

Let us denote:

\[
\phi _1\epsilon _1+\phi _2\epsilon _2=\phi _L\mbox {
and }\phi _3\epsilon _3+\phi _4\epsilon _4=\phi _R\mbox {.} 
\]

Hence from (\ref{bir}):

\begin{equation}
\begin{array}{c}
\mathbf{\Psi }\left( x_5,x_4\right) =\phi _L\left( 0,0\right) +\phi _R\left(
0,0\right) + \\ 
+\exp \left( -ih\left( nx_5+kx_4\right) \right) \left( \phi _L\left(
n,k\right) +\phi _R\left( n,k\right) \right) \mbox{.}
\end{array}
\label{bir1}
\end{equation}

If use denotation:

\[
\underline{\vartheta }=\left[ 
\begin{array}{cc}
\vartheta & 0_4 \\ 
0_4 & \vartheta
\end{array}
\right] \mbox {,} 
\]

\[
\gamma =\left[ 
\begin{array}{cc}
-\gamma _0 & 0_4 \\ 
0_4 & \gamma _0
\end{array}
\right] \mbox{, }\beta =\left[ 
\begin{array}{cc}
-\beta _4 & 0_4 \\ 
0_4 & \beta _4
\end{array}
\right] 
\]

and

\[
\underline{n}=\left[ 
\begin{array}{cc}
0_4 & 0_4 \\ 
0_4 & n1_4
\end{array}
\right] \mbox {, }\underline{k}=\left[ 
\begin{array}{cc}
0_4 & 0_4 \\ 
0_4 & k1_4
\end{array}
\right] 
\]

then the motion equation is the following:

\begin{equation}
\begin{array}{c}
\underline{\beta _0}i\partial _t\Psi +\underline{\beta _1}i\partial _x\Psi +%
\underline{\beta _2}i\partial _y\Psi +\underline{\beta _3}i\partial _z\Psi +
\\ 
-h\underline{n}\gamma \Psi -h\underline{k}\beta \Psi + \\ 
+\left( G_t\underline{\beta _0}+G_x\underline{\beta _1}+G_y\underline{\beta
_2}+G_z\underline{\beta _3}\right) \Psi + \\ 
+\left( K_t\underline{\beta _0}+K_x\underline{\beta _1}+K_y\underline{\beta
_2}+K_z\underline{\beta _3}\right) \underline{\gamma _5}\Psi =0
\end{array}
\label{mtn}
\end{equation}

and

\[
\left\{ 
\begin{array}{c}
\Psi ^{\dagger }\Psi =\rho \mbox{,} \\ 
\Psi ^{\dagger }\underline{\beta _1}\Psi =j_x\mbox{,} \\ 
\Psi ^{\dagger }\underline{\beta _2}\Psi =j_y\mbox{,} \\ 
\Psi ^{\dagger }\underline{\beta _3}\Psi =j_z\mbox{.}
\end{array}
\right| 
\]

If use the following denotation: $t=x_0,x=x_1,y=x_2,z=x_3,\partial _\mu
=\frac \partial {\partial _{x_\mu }}$ then the lagrangian has got the
following form:

\begin{equation}
\begin{array}{c}
\mathcal{L}_f=0.5i\left( \left( \sum_{\mu =0}^3\Psi ^{\dagger }\underline{%
\beta _\mu }\partial _\mu \Psi \right) -\left( \sum_{\mu =0}^3\partial _\mu
\Psi ^{\dagger }\underline{\beta _\mu }\Psi \right) \right) - \\ 
-\left( \Psi ^{\dagger }h\underline{n}\gamma \Psi +\Psi ^{\dagger }h%
\underline{k}\beta \Psi \right) + \\ 
+\Psi ^{\dagger }\left( \sum_{\mu =0}^3G_{x_\mu }\underline{\beta _\mu }%
\right) \Psi +\Psi ^{\dagger }\left( \sum_{\mu =0}^3K_{x_\mu }\underline{%
\beta _\mu }\right) \underline{\gamma _5}\Psi .
\end{array}
\label{lag}
\end{equation}

This lagrangian is invariant for the rotation of $xOy,yOz,xOz$ and for the
Lorentz transformation of $tOx,tOy,tOz$ and $G_{x_k}$ and $K_{x_k}$ behaves
as the 4-vector fields \cite{WGA}.

\subsubsection{Transformations}

If $U$ is an 8$\times $8 complex matrix, $\Psi ^{\prime }=U\Psi $ and

\begin{equation}
\left\{ 
\begin{array}{c}
\Psi ^{\prime \dagger }\underline{\beta _1}\Psi ^{\prime }=j_x\mbox{,} \\ 
\Psi ^{\prime \dagger }\underline{\beta _2}\Psi ^{\prime }=j_y\mbox{,} \\ 
\Psi ^{\prime \dagger }\underline{\beta _3}\Psi ^{\prime }=j_z
\end{array}
\right|  \label{uni}
\end{equation}

then for $1\leq k\leq 3$: $U^{\dagger }\underline{\beta _k}U=$\underline{$%
\beta _k$}. In this case a real numbers

$a",b",c",g",u",v",k,s,a`,b`,c`,g`,u`,v`,k`,s`$ exist for which:

\[
\underline{U}=\left[ 
\begin{array}{cccc}
\left( a"+b"i\right) 1_2 & 0_2 & \left( c"+ig"\right) 1_2 & 0_2 \\ 
0_2 & \left( a`+b`i\right) 1_2 & 0_2 & \left( c`+ig`\right) 1_2 \\ 
\left( u"+iv"\right) 1_2 & 0_2 & \left( k"+is"\right) 1_2 & 0_2 \\ 
0_2 & \left( u`+iv`\right) 1_2 & 0_2 & \left( k`+is`\right) 1_2
\end{array}
\right] . 
\]

If $\Psi ^{\prime \dagger }\Psi ^{\prime }=\rho $ then $U^{\dagger }U=1_8$ .
Hence:

\[
\begin{array}{c}
v"^2+b"^2+u"^2+a"^2=1, \\ 
c"^2+g"^2+k"^2+s"^2=1,
\end{array}
\]

\[
s"=-\frac{a"g"u"-u"b"c"+a"c"v"+b"g"v"}{u"^2+v"^2}, 
\]

\[
k"=\frac{-u"a"c"-u"b"g"+v"a"g"-b"c"v"}{u"^2+v"^2}. 
\]

\[
\begin{array}{c}
v`^2+b`^2+u`^2+a`^2=1, \\ 
c`^2+g`^2+k`^2+s`^2=1,
\end{array}
\]

\[
s`=-\frac{a`g`u`-u`b`c`+a`c`v`+b`g`v`}{u`^2+v`^2}, 
\]

\[
k`=\frac{-u`a`c`-u`b`g`+v`a`g`-b`c`v`}{u`^2+v`^2}. 
\]

$\underline{U}$ has got 4 eigenvalues: $\exp \left( i\alpha _1\right) $, $%
\exp \left( i\alpha _2\right) $, $\exp \left( i\alpha _3\right) $, $\exp
\left( i\alpha _4\right) $ for 8 orthogonal eigenvectors:

$\mathbf{\varepsilon }_{1,1}$, $\mathbf{\varepsilon }_{1,2}$, $\mathbf{%
\varepsilon }_{2,1}$, $\mathbf{\varepsilon }_{2,2}$, $\mathbf{\varepsilon }%
_{3,1}$, $\mathbf{\varepsilon }_{3,2}$, $\mathbf{\varepsilon }_{4,1}$, $%
\mathbf{\varepsilon }_{4,2}$.

Let

\[
K=\left[ 
\begin{array}{cccccccc}
\mathbf{\varepsilon }_{1,1} & \mathbf{\varepsilon }_{1,2} & \mathbf{%
\varepsilon }_{2,1} & \mathbf{\varepsilon }_{2,2} & \mathbf{\varepsilon }%
_{3,1} & \mathbf{\varepsilon }_{3,2} & \mathbf{\varepsilon }_{4,1} & \mathbf{%
\varepsilon }_{4,2}
\end{array}
\right] . 
\]

Let $\theta _1$, $\theta _2$, $\theta _3$, $\theta _4$ be the solution of
the system of the equations:

\[
\left\{ 
\begin{array}{c}
\theta _1+\theta _2+\theta _3+\theta _4=\alpha _1, \\ 
\theta _1+\theta _2-\theta _3-\theta _4=\alpha _1, \\ 
\theta _1-\theta _2+\theta _3-\theta _4=\alpha _1, \\ 
\theta _1-\theta _2-\theta _3+\theta _4=\alpha _1.
\end{array}
\right| 
\]

and

\[
U_1=\exp \left( i\theta _1\right) 1_8, 
\]

\[
U_2=K\left[ 
\begin{array}{cc}
\exp \left( i\theta _2\right) 1_4 & 0_4 \\ 
0_4 & \exp \left( -i\theta _2\right) 1_4
\end{array}
\right] K^{\dagger }, 
\]

\[
U_3=K\left[ 
\begin{array}{cccc}
\exp \left( i\theta _3\right) 1_2 & 0_2 & 0_2 & 0_2 \\ 
0_2 & \exp \left( -i\theta _3\right) 1_2 & 0_2 & 0_2 \\ 
0_2 & 0_2 & \exp \left( i\theta _3\right) 1_2 & 0_2 \\ 
0_2 & 0_2 & 0_2 & \exp \left( -i\theta _3\right) 1_2
\end{array}
\right] K^{\dagger }, 
\]

\[
U_4=K\left[ 
\begin{array}{cccc}
\exp \left( i\theta _4\right) 1_2 & 0_2 & 0_2 & 0_2 \\ 
0_2 & \exp \left( -i\theta _4\right) 1_2 & 0_2 & 0_2 \\ 
0_2 & 0_2 & \exp \left( -i\theta _4\right) 1_2 & 0_2 \\ 
0_2 & 0_2 & 0_2 & \exp \left( i\theta _4\right) 1_2
\end{array}
\right] K^{\dagger }. 
\]

In this case: 
\[
U_1U_2U_3U_4=U 
\]

and

\[
U_2=\left[ 
\begin{array}{cccc}
\exp \left( i\theta _2\right) 1_2 & 0_2 & 0_2 & 0_2 \\ 
0_2 & \exp \left( -i\theta _2\right) 1_2 & 0_2 & 0_2 \\ 
0_2 & 0_2 & \exp \left( i\theta _2\right) 1_2 & 0_2 \\ 
0_2 & 0_2 & 0_2 & \exp \left( -i\theta _2\right) 1_2
\end{array}
\right] 
\]

and a real number $a,b,c,g,u,v,k,s$ exist for which:

\[
U_3U_4=\left[ 
\begin{array}{cccc}
\left( a+ib\right) 1_2 & 0_2 & \left( c+ig\right) 1_2 & 0_2 \\ 
0_2 & \left( u+iv\right) 1_2 & 0_2 & \left( k+is\right) 1_2 \\ 
\left( -c+ig\right) 1_2 & 0_2 & \left( a-ib\right) 1_2 & 0_2 \\ 
0_2 & \left( -k+is\right) 1_2 & 0_2 & \left( u-iv\right) 1_2
\end{array}
\right] 
\]

and

\[
\begin{array}{c}
a^2+b^2+c^2+g^2=1, \\ 
u^2+v^2+r^2+s^2=1.
\end{array}
\]

If

\begin{equation}
U^{\left( +\right) }=\left[ 
\begin{array}{cccc}
1_2 & 0_2 & 0_2 & 0_2 \\ 
0_2 & \left( u+iv\right) 1_2 & 0_2 & \left( k+is\right) 1_2 \\ 
0_2 & 0_2 & 1_2 & 0_2 \\ 
0_2 & \left( -k+is\right) 1_2 & 0_2 & \left( u-iv\right) 1_2
\end{array}
\right]  \label{upls}
\end{equation}

and

\[
U^{\left( -\right) }=\left[ 
\begin{array}{cccc}
\left( a+ib\right) 1_2 & 0_2 & \left( c+ig\right) 1_2 & 0_2 \\ 
0_2 & 1_2 & 0_2 & 0_2 \\ 
\left( -c+ig\right) 1_2 & 0_2 & \left( a-ib\right) 1_2 & 0_2 \\ 
0_2 & 0_2 & 0_2 & 1_2
\end{array}
\right] 
\]

then

\[
U_3U_4=U^{\left( -\right) }U^{\left( +\right) }=U^{\left( +\right)
}U^{\left( -\right) }. 
\]

\paragraph{B-boson}

\[
U_1U_2=\left[ 
\begin{array}{cccc}
e^{i\left( \theta _1+\theta _2\right) } & 0 & 0 & 0 \\ 
0 & e^{i\left( \theta _1-\theta _2\right) } & 0 & 0 \\ 
0 & 0 & e^{i\left( \theta _1+\theta _2\right) } & 0 \\ 
0 & 0 & 0 & e^{i\left( \theta _1-\theta _2\right) }
\end{array}
\right] 
\]

Let $\chi $ and $\varsigma $ be the solution of the following set of
equations:

\[
\left\{ 
\begin{array}{c}
0.5\chi +\varsigma =\theta _1+\theta _2, \\ 
\chi +\varsigma =\theta _1-\theta _2\mbox{,}
\end{array}
\right| 
\]

i.e.:

\[
\begin{array}{c}
\chi =-4\theta _2\mbox{,} \\ 
\varsigma =\theta _1+3\theta _2\mbox{.}
\end{array}
\]

Let

\[
\overbrace{U}=\exp \left( i\varsigma \right) 1_8 
\]

and

\[
\widetilde{U}=\left[ 
\begin{array}{cccc}
\exp \left( i\frac \chi 2\right) 1_2 & 0_2 & 0_2 & 0_2 \\ 
0_2 & \exp \left( i\chi \right) 1_2 & 0_2 & 0_2 \\ 
0_2 & 0_2 & \exp \left( i\frac \chi 2\right) 1_2 & 0_2 \\ 
0_2 & 0_2 & 0_2 & \exp \left( i\chi \right) 1_2
\end{array}
\right] \mbox{.} 
\]

In that case:

\[
\widetilde{U}\overbrace{U}=U_1U_2\mbox{.} 
\]

Let $g_1$ be a positive real number and for $\mu \in \left\{ t,x,y,z\right\} 
$: $F_\mu $ and $B_\mu $ be the solutions of the following systems of the
equations:

\[
\left\{ 
\begin{array}{c}
-0.5g_1B_\mu +F_\mu =G_\mu +K_\mu \\ 
-g_1B_\mu +F_\mu =G_\mu -K_\mu
\end{array}
\right| 
\]

i.e.:

\begin{eqnarray*}
B_\mu &=&\frac 4{g_1}K_\mu \\
F_\mu &=&G_\mu +3K_\mu \mbox{.}
\end{eqnarray*}

Let {\it the charge matrix} be defined as the following:

\[
\underline{Y}=-\left[ 
\begin{array}{cccc}
1_2 & 0_2 & 0_2 & 0_2 \\ 
0_2 & 2\cdot 1_2 & 0_2 & 0_2 \\ 
0_2 & 0_2 & 1_2 & 0_2 \\ 
0_2 & 0_2 & 0_2 & 2\cdot 1_2
\end{array}
\right] 
\]

In that case from (\ref{mtn}):

\begin{equation}
\begin{array}{c}
\underline{\beta _0}i\partial _t\Psi +\underline{\beta _1}i\partial _x\Psi +%
\underline{\beta _2}i\partial _y\Psi +\underline{\beta _3}i\partial _z\Psi +
\\ 
-h\underline{n}\gamma \Psi -h\underline{k}\beta \Psi + \\ 
+\left( F_t\underline{\beta _0}+F_x\underline{\beta _1}+F_y\underline{\beta
_2}+F_z\underline{\beta _3}\right) \Psi + \\ 
+0.5g_1\underline{Y}\left( B_t\underline{\beta _0}+B_x\underline{\beta _1}%
+B_y\underline{\beta _2}+B_z\underline{\beta _3}\right) \Psi =0\mbox{.}
\end{array}
\label{umt1}
\end{equation}

Let

\[
\begin{array}{c}
\Psi \rightarrow \Psi `=\left( \widetilde{U}\Psi \right) \mbox{,} \\ 
n\rightarrow n`\mbox{,} \\ 
k\rightarrow k`\mbox{,} \\ 
F_\mu \rightarrow F_\mu `\mbox{,} \\ 
B_\mu \rightarrow B_\mu `
\end{array}
\]

then:

\[
\begin{array}{c}
\underline{\beta _0}i\partial _t\left( \widetilde{U}\Psi \right) +\underline{%
\beta _1}i\partial _x\left( \widetilde{U}\Psi \right) +\underline{\beta _2}%
i\partial _y\left( \widetilde{U}\Psi \right) +\underline{\beta _3}i\partial
_z\left( \widetilde{U}\Psi \right) + \\ 
-h\underline{n}`\gamma \widetilde{U}\Psi -h\underline{k}`\beta \widetilde{U}%
\Psi + \\ 
+\left( F_t\underline{`\beta _0}+F_x`\underline{\beta _1}+F_y`\underline{%
\beta _2}+F_z`\underline{\beta _3}\right) \widetilde{U}\Psi + \\ 
+0.5g_1\underline{Y}\left( B_t`\underline{\beta _0}+B_x`\underline{\beta _1}%
+B_y`\underline{\beta _2}+B_z`\underline{\beta _3}\right) \widetilde{U}\Psi
=0
\end{array}
\]

hence:

\[
\begin{array}{c}
\underline{\beta _0}i\left( \partial _t\widetilde{U}\right) \Psi +\underline{%
\beta _0}i\widetilde{U}\partial _t\Psi +\underline{\beta _1}i\left( \partial
_x\widetilde{U}\right) \Psi +\underline{\beta _1}i\widetilde{U}\partial
_x\Psi + \\ 
+\underline{\beta _2}i\left( \partial _y\widetilde{U}\right) \Psi +%
\underline{\beta _2}i\widetilde{U}\partial _y\Psi +\underline{\beta _3}%
i\left( \partial _z\widetilde{U}\right) \Psi +\underline{\beta _3}i%
\widetilde{U}\partial _z\Psi + \\ 
-h\underline{n}`\gamma \widetilde{U}\Psi -h\underline{k}`\beta \widetilde{U}%
\Psi + \\ 
+\left( F_t\underline{`\beta _0}+F_x`\underline{\beta _1}+F_y`\underline{%
\beta _2}+F_z`\underline{\beta _3}\right) \widetilde{U}\Psi + \\ 
+0.5g_1\underline{Y}\left( B_t`\underline{\beta _0}+B_x`\underline{\beta _1}%
+B_y`\underline{\beta _2}+B_z`\underline{\beta _3}\right) \widetilde{U}\Psi =
\\ 
=0\mbox{.}
\end{array}
\]

Since

\[
\partial _\mu \widetilde{U}=i\frac{\partial _\mu \chi }2\left[ 
\begin{array}{cccc}
\exp \left( i\frac \chi 2\right) & 0 & 0 & 0 \\ 
0 & 2\exp \left( i\chi \right) & 0 & 0 \\ 
0 & 0 & \exp \left( i\frac \chi 2\right) & 0 \\ 
0 & 0 & 0 & 2\exp \left( i\chi \right)
\end{array}
\right] 
\]

then

\[
\partial _\mu \widetilde{U}=-i\frac{\partial _\mu \chi }2\underline{Y}%
\widetilde{U}\mbox{;} 
\]

Hence

\[
\begin{array}{c}
\left( \underline{\beta _0}i\widetilde{U}\partial _t+\underline{\beta _1}i%
\widetilde{U}\partial _x+\underline{\beta _2}i\widetilde{U}\partial _y+%
\underline{\beta _3}i\widetilde{U}\partial _z\right) \Psi + \\ 
-h\left( \underline{n}`\gamma \widetilde{U}+\underline{k}`\beta \widetilde{U}%
\right) \Psi + \\ 
+\left( F_t\underline{`\beta _0}+F_x`\underline{\beta _1}+F_y`\underline{%
\beta _2}+F_z`\underline{\beta _3}\right) \widetilde{U}\Psi + \\ 
+0.5\left( 
\begin{array}{c}
\left( g_1\underline{Y}\underline{\beta _0}B_t`+\underline{\beta _0}%
\underline{Y}\partial _t\chi \right) +\left( g_1\underline{Y}\underline{%
\beta _1}B_x`+\underline{\beta _1}\underline{Y}\partial _x\chi \right) + \\ 
+\left( g_1\underline{Y}\underline{\beta _2}B_y`+\underline{\beta _2}%
\underline{Y}\partial _y\chi \right) +\left( g_1\underline{Y}\underline{%
\beta _3}B_z`+\underline{\beta _3}\underline{Y}\partial _z\chi \right)
\end{array}
\right) \widetilde{U}\Psi = \\ 
=0
\end{array}
\]

Since $\underline{Y}\underline{\beta _\mu }=\underline{\beta _\mu }%
\underline{Y}$ then

\[
\begin{array}{c}
\left( \underline{\beta _0}i\widetilde{U}\partial _t+\underline{\beta _1}i%
\widetilde{U}\partial _x+\underline{\beta _2}i\widetilde{U}\partial _y+%
\underline{\beta _3}i\widetilde{U}\partial _z\right) \Psi + \\ 
-h\left( \underline{n}`\gamma \widetilde{U}+\underline{k}`\beta \widetilde{U}%
\right) \Psi + \\ 
+\left( F_t\underline{`\beta _0}+F_x`\underline{\beta _1}+F_y`\underline{%
\beta _2}+F_z`\underline{\beta _3}\right) \widetilde{U}\Psi + \\ 
+0.5\underline{Y}\left( 
\begin{array}{c}
\underline{\beta _0}\left( g_1B_t`+\partial _t\chi \right) +\underline{\beta
_1}\left( g_1B_x`+\partial _x\chi \right) + \\ 
+\underline{\beta _2}\left( g_1B_y`+\partial _y\chi \right) +\underline{%
\beta _3}\left( g_1B_z`+\partial _z\chi \right)
\end{array}
\right) \widetilde{U}\Psi = \\ 
=0
\end{array}
\]

hence:

\[
\begin{array}{c}
\widetilde{U}^{\dagger }\left( \underline{\beta _0}i\widetilde{U}\partial _t+%
\underline{\beta _1}i\widetilde{U}\partial _x+\underline{\beta _2}i%
\widetilde{U}\partial _y+\underline{\beta _3}i\widetilde{U}\partial
_z\right) \Psi + \\ 
-h\widetilde{U}^{\dagger }\left( \underline{n}`\gamma \widetilde{U}+%
\underline{k}`\beta \widetilde{U}\right) \Psi + \\ 
+\widetilde{U}^{\dagger }\left( F_t\underline{`\beta _0}+F_x`\underline{%
\beta _1}+F_y`\underline{\beta _2}+F_z`\underline{\beta _3}\right) 
\widetilde{U}\Psi + \\ 
+0.5\widetilde{U}^{\dagger }\underline{Y}\left( 
\begin{array}{c}
\underline{\beta _0}\left( g_1B_t`+\partial _t\chi \right) +\underline{\beta
_1}\left( g_1B_x`+\partial _x\chi \right) + \\ 
+\underline{\beta _2}\left( g_1B_y`+\partial _y\chi \right) +\underline{%
\beta _3}\left( g_1B_z`+\partial _z\chi \right)
\end{array}
\right) \widetilde{U}\Psi = \\ 
=0\mbox{.}
\end{array}
\]

Because:

\[
\begin{array}{c}
\widetilde{U}^{\dagger }\gamma \widetilde{U}=\cos \left( \frac \chi 2\right)
\gamma -\sin \left( \frac \chi 2\right) \beta \mbox{,} \\ 
\widetilde{U}^{\dagger }\beta \widetilde{U}=\cos \left( \frac \chi 2\right)
\beta +\sin \left( \frac \chi 2\right) \gamma \mbox{,} \\ 
\widetilde{U}^{\dagger }\widetilde{U}=1_8\mbox{,} \\ 
\underline{\beta _\mu }\widetilde{U}=\widetilde{U}\underline{\beta _\mu }%
\mbox{,} \\ 
U^{\dagger }\underline{Y}U=\underline{Y}
\end{array}
\]

then

\[
\begin{array}{c}
\left( \underline{\beta _0}i\partial _t+\underline{\beta _1}i\partial _x+%
\underline{\beta _2}i\partial _y+\underline{\beta _3}i\partial _z\right)
\Psi + \\ 
-h\left( \underline{n}`\left( \cos \left( \frac \chi 2\right) \gamma -\sin
\left( \frac \chi 2\right) \beta \right) +\underline{k}`\left( \cos \left(
\frac \chi 2\right) \beta +\sin \left( \frac \chi 2\right) \gamma \right)
\right) \Psi + \\ 
+\left( F_t\underline{`\beta _0}+F_x`\underline{\beta _1}+F_y`\underline{%
\beta _2}+F_z`\underline{\beta _3}\right) \Psi + \\ 
+0.5\underline{Y}\left( 
\begin{array}{c}
\underline{\beta _0}\left( g_1B_t`+\partial _t\chi \right) +\underline{\beta
_1}\left( g_1B_x`+\partial _x\chi \right) + \\ 
+\underline{\beta _2}\left( g_1B_y`+\partial _y\chi \right) +\underline{%
\beta _3}\left( g_1B_z`+\partial _z\chi \right)
\end{array}
\right) \Psi = \\ 
=0\mbox{.}
\end{array}
\]

Therefore from (\ref{umt1}):

\begin{equation}
\begin{array}{c}
F_x`=F_x\mbox{,} \\ 
B_\mu `=B_\mu -\frac 1{g_1}\partial _\mu \chi \mbox{,} \\ 
n`=-k\sin \frac \chi 2+n\cos \frac \chi 2\mbox{,} \\ 
k`=k\cos \frac \chi 2+n\sin \frac \chi 2\mbox{.}
\end{array}
\label{tt4}
\end{equation}

But $k$ and $n$ are an integer numbers and $k`$ and $n`$ must be an integer
numbers, too.

A triplet $\left\langle l,n,k\right\rangle $ of integer numbers is {\it a
Fermat triplet} if

\[
l^2=n^2+k^2\mbox{.} 
\]

Let $\varepsilon $ be any tiny positive real number. An integer number $l$
is {\it a father number with a precise }$\varepsilon $ if for each real
number $\chi $ and for every Fermat triplet $\left\langle l,n,k\right\rangle 
$ a Fermat triplet $\left\langle l,n`,k`\right\rangle $ exists for which:

\[
\begin{array}{c}
\left| -k\sin \frac \chi 2+n\cos \frac \chi 2-n`\right| <\varepsilon \mbox{,}
\\ 
\left| k\cos \frac \chi 2+n\sin \frac \chi 2-k`\right| <\varepsilon \mbox{.}
\end{array}
\]

{\it For every }$\varepsilon ${\it : denumerable many of a father
numbers with a precise }$\varepsilon ${\it \ exist.}

Excuse me, but I mean that a masses of the real members of the particles
families are defined by a father numbers with a precise $h$. I.e.
denumerable many of a families exist.

Therefore for the (\ref{tt4}) transformation from (\ref{bir1}):

\[
\begin{array}{c}
\mathbf{\Psi }\left( x_5,x_4\right) =\sum_{r=1}^4\phi _r\left( 0,0\right)
\epsilon _r+\exp \left( -ih\left( nx_5+kx_4\right) \right) \sum_{k=1}^4\phi
_k\left( n,k\right) \epsilon _k= \\ 
=\phi _L\left( 0,0\right) +\phi _R\left( 0,0\right) +\exp \left( -ih\left(
nx_5+kx_4\right) \right) \left( \phi _L\left( n,k\right) +\phi _R\left(
n,k\right) \right) \rightarrow
\end{array}
\]

\[
\begin{array}{c}
\rightarrow \mathbf{\Psi }`\left( x_5,x_4\right) = \\ 
=\exp \left( i\frac \chi 2\right) \phi _L\left( 0,0\right) +\exp \left(
i\chi \right) \phi _R\left( 0,0\right) + \\ 
+\exp \left( -ih\left( \left( -k\sin \frac \chi 2+n\cos \frac \chi 2\right)
x_5+\left( k\cos \frac \chi 2+n\sin \frac \chi 2\right) x_4\right) \right)
\cdot \\ 
\cdot \left( \exp \left( i\frac \chi 2\right) \phi _L\left( n,k\right) +\exp
\left( i\chi \right) \phi _R\left( n,k\right) \right) \mbox {.}
\end{array}
\]

\paragraph{$U^{\left( -\right) }$ transformation}

$U^{\left( -\right) }$ has got the following eigenvalues and eigenvectors:

for the eigenvalue $1$: eigenvectors:

\begin{equation}
\underline{\iota }_1=\left[ 
\begin{array}{c}
0 \\ 
0 \\ 
1 \\ 
0 \\ 
0 \\ 
0 \\ 
0 \\ 
0
\end{array}
\right] ,\underline{\iota }_2=\left[ 
\begin{array}{c}
0 \\ 
0 \\ 
0 \\ 
1 \\ 
0 \\ 
0 \\ 
0 \\ 
0
\end{array}
\right] ,\underline{\iota }_5=\left[ 
\begin{array}{c}
0 \\ 
0 \\ 
0 \\ 
0 \\ 
0 \\ 
0 \\ 
1 \\ 
0
\end{array}
\right] ,\underline{\iota }_6=\left[ 
\begin{array}{c}
0 \\ 
0 \\ 
0 \\ 
0 \\ 
0 \\ 
0 \\ 
0 \\ 
1
\end{array}
\right] ;  \label{si1}
\end{equation}

for the eigenvalue $w=a+i\sqrt{1-a^2}$: eigenvectors:

\begin{equation}
\underline{\iota }_3=\frac 1{\sqrt{2}\sqrt{\sqrt{1-a^2}\left( b+\sqrt{1-a^2}%
\right) }}\left[ 
\begin{array}{c}
b+\sqrt{1-a^2} \\ 
0 \\ 
0 \\ 
0 \\ 
ic+g \\ 
0 \\ 
0 \\ 
0
\end{array}
\right] ,  \label{si2}
\end{equation}

\begin{equation}
\underline{\iota }_4=\frac 1{\sqrt{2}\sqrt{\sqrt{1-a^2}\left( b+\sqrt{1-a^2}%
\right) }}\left[ 
\begin{array}{c}
0 \\ 
b+\sqrt{1-a^2} \\ 
0 \\ 
0 \\ 
0 \\ 
ic+g \\ 
0 \\ 
0
\end{array}
\right] \mbox{;}  \label{si3}
\end{equation}

for eigenvalue $w^{*}=a-i\sqrt{1-a^2}$: eigenvectors:

\begin{equation}
\underline{\iota }_7=\frac 1{\sqrt{2}\sqrt{\sqrt{1-a^2}\left( b+\sqrt{1-a^2}%
\right) }}\left[ 
\begin{array}{c}
ic-g \\ 
0 \\ 
0 \\ 
0 \\ 
b+\sqrt{1-a^2} \\ 
0 \\ 
0 \\ 
0
\end{array}
\right] \mbox{,}  \label{si4}
\end{equation}

\begin{equation}
\underline{\iota }_8=\frac 1{\sqrt{2}\sqrt{\sqrt{1-a^2}\left( b+\sqrt{1-a^2}%
\right) }}\left[ 
\begin{array}{c}
0 \\ 
ic-g \\ 
0 \\ 
0 \\ 
0 \\ 
b+\sqrt{1-a^2} \\ 
0 \\ 
0
\end{array}
\right] \mbox{.}  \label{si5}
\end{equation}

Hence the space of $U^{\left( -\right) }$ is divided on three orthogonal
subspace:

the 4-dimensional $\mathcal{U}_1^{\left( -\right) }$on the basis $%
\left\langle \underline{\iota }_1,\underline{\iota }_2,\underline{\iota }_5,%
\underline{\iota }_6\right\rangle $ with eigenvalue $1$,

the 2-dimensional $\mathcal{U}_w^{\left( -\right) }$on the basis $%
\left\langle \underline{\iota }_3,\underline{\iota }_4\right\rangle $ with
eigenvalue $w$ and

the 2-dimensional $\mathcal{U}_{w^{*}}^{\left( -\right) }$on the basis $%
\left\langle \underline{\iota }_7,\underline{\iota }_8\right\rangle $ with
eigenvalue $w^{*}$.

Let

\[
\widehat{\underline{\iota }}_k=\gamma \underline{\iota }_k\mbox{.} 
\]

In this case $\left\langle \widehat{\underline{\iota }}_3,\widehat{%
\underline{\iota }}_4,\widehat{\underline{\iota }}_7,\widehat{\underline{%
\iota }}_8\right\rangle $ is the orthonormal basis of $\mathcal{U}_1^{\left(
-\right) }$.

Let $\mathcal{U}_{\circ }^{\left( -\right) }$ be the space on the basis $%
\left\langle \widehat{\underline{\iota }}_3,\widehat{\underline{\iota }}_4,%
\underline{\iota }_3,\underline{\iota }_4\right\rangle $ and $\mathcal{U}%
_{*}^{\left( -\right) }$ be the space on the basis $\left\langle \widehat{%
\underline{\iota }}_7,\widehat{\underline{\iota }}_8,\underline{\iota }_7,%
\underline{\iota }_8\right\rangle $.

\begin{equation}
\begin{array}{c}
\Psi _{\circ }=\pi _{\circ }\Psi \mbox{, }\Psi _{*}=\pi _{*}\Psi \mbox{,} \\ 
\Psi _{\circ }\in \mathcal{U}_{\circ }^{\left( -\right) }\mbox{ and }\Psi
_{*}\in \mathcal{U}_{*}^{\left( -\right) }\mbox{.}
\end{array}
\label{ux}
\end{equation}

In this case:

\[
\pi _{\circ }=\frac 1{2\sqrt{1-a^2}}\left[ 
\begin{array}{cc}
\left( b+\sqrt{1-a^2}\right) 1_4 & \left( -ic+g\right) \gamma _5 \\ 
\left( ic+g\right) \gamma _5 & \left( \sqrt{1-a^2}-b\right) 1_4
\end{array}
\right] \mbox{,} 
\]

\[
\pi _{*}=\frac 1{2\sqrt{1-a^2}}\left[ 
\begin{array}{cc}
\left( \sqrt{1-a^2}-b\right) 1_4 & \left( ic-g\right) \gamma _5 \\ 
\left( -g-ic\right) \gamma _5 & \left( b+\sqrt{1-a^2}\right) 1_4
\end{array}
\right] \mbox{.} 
\]

Hence

\begin{equation}
\begin{array}{c}
\mathbf{\Psi }_{\circ }\left( x_5,x_4\right) =\frac 1{2\sqrt{1-a^2}}\cdot \\ 
\cdot \left( 
\begin{array}{c}
\left( b+\sqrt{\left( 1-a^2\right) }\right) \left( \phi _L\left( 0,0\right)
+\phi _R\left( 0,0\right) \right) - \\ 
-\left( ic-g\right) \left( \phi _L\left( n,k\right) -\phi _R\left(
n,k\right) \right) + \\ 
+\exp \left( -ih\left( nx_5+kx_4\right) \right) \cdot \\ 
\cdot \left( 
\begin{array}{c}
\left( ic+g\right) \left( \phi _L\left( 0,0\right) -\phi _R\left( 0,0\right)
\right) + \\ 
+\left( b+\sqrt{\left( 1-a^2\right) }\right) \left( \phi _L\left( n,k\right)
+\phi _R\left( n,k\right) \right)
\end{array}
\right)
\end{array}
\right) \mbox{,}
\end{array}
\label{fr1}
\end{equation}

\begin{equation}
\begin{array}{c}
\mathbf{\Psi }_{*}\left( x_5,x_4\right) =\frac 1{2\sqrt{1-a^2}}\cdot \\ 
\cdot \left( 
\begin{array}{c}
\left( \sqrt{\left( 1-a^2\right) }-b\right) \left( \phi _L\left( 0,0\right)
+\phi _R\left( 0,0\right) \right) + \\ 
+\left( ic-g\right) \left( \phi _L\left( n,k\right) -\phi _R\left(
n,k\right) \right) + \\ 
+\exp \left( -ih\left( nx_5+kx_4\right) \right) \cdot \\ 
\cdot \left( 
\begin{array}{c}
\left( g+ic\right) \left( -\phi _L\left( 0,0\right) +\phi _R\left(
0,0\right) \right) + \\ 
+\left( b+\sqrt{\left( 1-a^2\right) }\right) \left( \phi _L\left( n,k\right)
+\phi _R\left( n,k\right) \right)
\end{array}
\right)
\end{array}
\right) \mbox{.}
\end{array}
\label{fr2}
\end{equation}

If $\lambda $ is the angle of the $U^{\left( -\right) }$ eigenvalue (i.e. $%
w=a+i\sqrt{1-a^2}$ and $\cos \lambda =a$ and $\sin \lambda =\sqrt{1-a^2}$ )
then

\begin{equation}
\begin{array}{c}
U^{\left( -\right) \dagger }\gamma U^{\left( -\right) }=\left( \gamma \cos
\lambda +\sin \lambda \left( \pi _{\circ }-\pi _{*}\right) \beta \right) %
\mbox{,} \\ 
U^{\left( -\right) \dagger }\beta U^{\left( -\right) }=\left( \beta \cos
\lambda -\sin \lambda \left( \pi _{\circ }-\pi _{*}\right) \gamma \right) %
\mbox{.}
\end{array}
\label{gb}
\end{equation}

Let

\begin{equation}
\begin{array}{c}
\underline{n}\rightarrow \underline{n}`=\left( \underline{n}\cos \lambda +%
\underline{k}\sin \lambda \left( \pi _{\circ }-\pi _{*}\right) \right) %
\mbox{,} \\ 
\underline{k}\rightarrow \underline{k}`=\left( \underline{k}\cos \lambda -%
\underline{n}\sin \lambda \left( \pi _{\circ }-\pi _{*}\right) \right) %
\mbox{,} \\ 
\Psi \rightarrow \Psi `=U^{\left( -\right) }\Psi \mbox{,} \\ 
F_\mu \rightarrow F_\mu `=F_\mu \mbox{,} \\ 
B_\mu \rightarrow B_\mu `=B_\mu
\end{array}
\label{tt}
\end{equation}

and the motion equation for $\Psi `$ be (\ref{umt1}):

\begin{equation}
\begin{array}{c}
\sum_{\mu =0}^3\underline{\beta _\mu }i\partial _\mu \Psi `-h\left( 
\underline{n}`\gamma +\underline{k}`\beta \right) \Psi `+\sum_{\mu
=0}^3F_\mu `\underline{\beta _\mu }\Psi `+ \\ 
+0.5g_1\underline{Y}\sum_{\mu =0}^3B_\mu `\underline{\beta _\mu }\Psi
`=S\Psi `\mbox{.}
\end{array}
\label{um2}
\end{equation}

Hence:

\[
\begin{array}{c}
\sum_{\mu =0}^3\underline{\beta _\mu }i\left( \partial _\mu U^{\left(
-\right) }\right) \Psi +\sum_{\mu =0}^3\underline{\beta _\mu }iU^{\left(
-\right) }\left( \partial _\mu \Psi \right) - \\ 
-h\left( \underline{n}`\gamma +\underline{k}`\beta \right) \left( U^{\left(
-\right) }\Psi \right) +\sum_{\mu =0}^3F_\mu `\underline{\beta _\mu }\left(
U^{\left( -\right) }\Psi \right) + \\ 
+0.5g_1\underline{Y}\sum_{\mu =0}^3B_\mu `\underline{\beta _\mu }\left(
U^{\left( -\right) }\Psi \right) =S\left( U^{\left( -\right) }\Psi \right) %
\mbox{,}
\end{array}
\]

then

\[
\begin{array}{c}
U^{\left( -\right) \dagger }\sum_{\mu =0}^3\underline{\beta _\mu }i\left(
\partial _\mu U^{\left( -\right) }\right) \Psi +U^{\left( -\right) \dagger
}\sum_{\mu =0}^3\underline{\beta _\mu }iU^{\left( -\right) }\left( \partial
_\mu \Psi \right) - \\ 
-U^{\left( -\right) \dagger }h\left( 
\begin{array}{c}
\left( \underline{n}\cos \lambda +\underline{k}\sin \lambda \left( \pi
_{\circ }-\pi _{*}\right) \right) \gamma + \\ 
+\left( \underline{k}\cos \lambda -\underline{n}\sin \lambda \left( \pi
_{\circ }-\pi _{*}\right) \right) \beta
\end{array}
\right) U^{\left( -\right) }\Psi + \\ 
+U^{\left( -\right) \dagger }\sum_{\mu =0}^3F_\mu \underline{\beta _\mu }%
U^{\left( -\right) }\Psi ++U^{\left( -\right) \dagger }0.5g_1\underline{Y}%
\sum_{\mu =0}^3B_\mu \underline{\beta _\mu }U^{\left( -\right) }\Psi = \\ 
=U^{\left( -\right) \dagger }SU^{\left( -\right) }\Psi \mbox{.}
\end{array}
\]

Since

\[
\begin{array}{c}
U^{\left( -\right) \dagger }\underline{\beta _\mu }=\underline{\beta _\mu }%
U^{\left( -\right) \dagger }\mbox{,} \\ 
\underline{Y}U^{\left( -\right) }=U^{\left( -\right) }\underline{Y}\mbox{,}
\\ 
U^{\left( -\right) \dagger }\left( \pi _{\circ }-\pi _{*}\right) =\left( \pi
_{\circ }-\pi _{*}\right) U^{\left( -\right) \dagger }
\end{array}
\]

then

\[
\begin{array}{c}
\sum_{\mu =0}^3\underline{\beta _\mu }iU^{\left( -\right) \dagger }\left(
\partial _\mu U^{\left( -\right) }\right) \Psi +\sum_{\mu =0}^3\underline{%
\beta _\mu }i\partial _\mu \Psi - \\ 
-h\left( 
\begin{array}{c}
\left( \underline{n}\cos \lambda +\underline{k}\sin \lambda \left( \pi
_{\circ }-\pi _{*}\right) \right) U^{\left( -\right) \dagger }\gamma
U^{\left( -\right) }+ \\ 
+\left( \underline{k}\cos \lambda -\underline{n}\sin \lambda \left( \pi
_{\circ }-\pi _{*}\right) \right) U^{\left( -\right) \dagger }\beta
U^{\left( -\right) }
\end{array}
\right) \Psi + \\ 
+\sum_{\mu =0}^3F_\mu \underline{\beta _\mu }\Psi +0.5g_1\underline{Y}%
\sum_{\mu =0}^3B_\mu \underline{\beta _\mu }\Psi = \\ 
=U^{\left( -\right) \dagger }SU^{\left( -\right) }\Psi \mbox{.}
\end{array}
\]

From (\ref{gb}):

\[
\begin{array}{c}
\sum_{\mu =0}^3\underline{\beta _\mu }iU^{\left( -\right) \dagger }\left(
\partial _\mu U^{\left( -\right) }\right) \Psi +\sum_{\mu =0}^3\underline{%
\beta _\mu }i\partial _\mu \Psi - \\ 
-h\left( 
\begin{array}{c}
\left( \underline{n}\cos \lambda +\underline{k}\sin \lambda \left( \pi
_{\circ }-\pi _{*}\right) \right) \left( \gamma \cos \lambda +\sin \lambda
\left( \pi _{\circ }-\pi _{*}\right) \beta \right) + \\ 
+\left( \underline{k}\cos \lambda -\underline{n}\sin \lambda \left( \pi
_{\circ }-\pi _{*}\right) \right) \left( \beta \cos \lambda -\sin \lambda
\left( \pi _{\circ }-\pi _{*}\right) \gamma \right)
\end{array}
\right) \Psi + \\ 
+\sum_{\mu =0}^3F_\mu \underline{\beta _\mu }\Psi +0.5g_1\underline{Y}%
\sum_{\mu =0}^3B_\mu \underline{\beta _\mu }\Psi = \\ 
=U^{\left( -\right) \dagger }SU^{\left( -\right) }\Psi \mbox{.}
\end{array}
\]

Since

\[
\left( \pi _{\circ }-\pi _{*}\right) \left( \pi _{\circ }-\pi _{*}\right)
=1_8 
\]

then

\[
\begin{array}{c}
\sum_{\mu =0}^3\underline{\beta _\mu }iU^{\left( -\right) \dagger }\left(
\partial _\mu U^{\left( -\right) }\right) \Psi + \\ 
+\sum_{\mu =0}^3\underline{\beta _\mu }i\partial _\mu \Psi -h\left( 
\underline{n}\gamma +\underline{k}\beta \right) \Psi +\sum_{\mu =0}^3F_\mu 
\underline{\beta _\mu }\Psi +0.5g_1\underline{Y}\sum_{\mu =0}^3B_\mu 
\underline{\beta _\mu }\Psi = \\ 
=U^{\left( -\right) \dagger }SU^{\left( -\right) }\Psi \mbox{.}
\end{array}
\]

Hence from (\ref{umt1}):

\[
\sum_{\mu =0}^3\underline{\beta _\mu }iU^{\left( -\right) \dagger }\left(
\partial _\mu U^{\left( -\right) }\right) =U^{\left( -\right) \dagger
}SU^{\left( -\right) } 
\]

and

\[
S=\sum_{\mu =0}^3\underline{\beta _\mu }i\left( \partial _\mu U^{\left(
-\right) }\right) U^{\left( -\right) \dagger } 
\]

Therefore from (\ref{um2}) the motion equation for the transformation (\ref
{tt}) is the following:

\begin{equation}
\begin{array}{c}
\sum_{\mu =0}^3\underline{\beta _\mu }i\partial _\mu \Psi `-\sum_{\mu =0}^3%
\underline{\beta _\mu }i\left( \partial _\mu U^{\left( -\right) }\right)
U^{\left( -\right) \dagger }\Psi `- \\ 
-h\left( \underline{n}`\gamma +\underline{k}`\beta \right) \Psi `+ \\ 
+\sum_{\mu =0}^3F_\mu \underline{\beta _\mu }\Psi `+0.5g_1\underline{Y}%
\sum_{\mu =0}^3B_\mu \underline{\beta _\mu }\Psi `=0\mbox{.}
\end{array}
\label{um3}
\end{equation}

\subparagraph{$W$-bosons}

Let $g_2$ be a positive real number.

If design ($a,b,c,g$ form $U^{\left( -\right) }$):

\[
\begin{array}{c}
W_{0,\mu }=-2\frac 1{g_2g}\left( 
\begin{array}{c}
g\left( \partial _\mu a\right) b-g\left( \partial _\mu b\right) a+\left(
\partial _\mu c\right) g^2+ \\ 
+a\left( \partial _\mu a\right) c+b\left( \partial _\mu b\right) c+c^2\left(
\partial _\mu c\right)
\end{array}
\right) \\ 
W_{1,\mu }=-2\frac 1{g_2g}\left( 
\begin{array}{c}
\left( \partial _\mu a\right) a^2-bg\left( \partial _\mu c\right) +a\left(
\partial _\mu b\right) b+ \\ 
+a\left( \partial _\mu c\right) c+g^2\left( \partial _\mu a\right) +c\left(
\partial _\mu b\right) g
\end{array}
\right) \\ 
W_{2,\mu }=-2\frac 1{g_2g}\left( 
\begin{array}{c}
g\left( \partial _\mu a\right) c-a\left( \partial _\mu a\right) b-b^2\left(
\partial _\mu b\right) - \\ 
-c\left( \partial _\mu c\right) b-\left( \partial _\mu b\right) g^2-\left(
\partial _\mu c\right) ga
\end{array}
\right)
\end{array}
\]

and

\[
W_\mu =\left[ 
\begin{array}{cccc}
W_{0,\mu }1_2 & 0_2 & \left( W_{1,\mu }-iW_{2,\mu }\right) 1_2 & 0_2 \\ 
0_2 & 0_2 & 0_2 & 0_2 \\ 
\left( W_{1,\mu }+iW_{2,\mu }\right) 1_2 & 0_2 & -W_{0,\mu }1_2 & 0_2 \\ 
0_2 & 0_2 & 0_2 & 0_2
\end{array}
\right] 
\]

then

\begin{equation}
-i\left( \partial _\mu U^{\left( -\right) }\right) U^{\left( -\right)
\dagger }=\frac 12g_2W_\mu  \label{w}
\end{equation}

and from (\ref{um3}):

\begin{equation}
\begin{array}{c}
\sum_{\mu =0}^3\underline{\beta _\mu }i\left( \partial _\mu -i\frac
12g_2W_\mu \right) \Psi `- \\ 
-h\left( \underline{n}`\gamma +\underline{k}`\beta \right) \Psi `+ \\ 
+\sum_{\mu =0}^3F_\mu \underline{\beta _\mu }\Psi `+0.5g_1\underline{Y}%
\sum_{\mu =0}^3B_\mu \underline{\beta _\mu }\Psi `=0\mbox{.}
\end{array}
\label{hW}
\end{equation}

Let

\[
\stackrel{,}{U}=\left[ 
\begin{array}{cccc}
\left( \stackrel{,}{a}+i\stackrel{,}{b}\right) 1_2 & 0_2 & \left( \stackrel{,%
}{c}+i\stackrel{,}{g}\right) 1_2 & 0_2 \\ 
0_2 & 1_2 & 0_2 & 0_2 \\ 
\left( -\stackrel{,}{c}+i\stackrel{,}{g}\right) 1_2 & 0_2 & \left( \stackrel{%
,}{a}-i\stackrel{,}{b}\right) 1_2 & 0_2 \\ 
0_2 & 0_2 & 0_2 & 1_2
\end{array}
\right] \mbox{,} 
\]

\[
\stackrel{,}{\pi }_{\circ }=\frac 1{2\sqrt{1-\stackrel{,}{a}^2}}\left[ 
\begin{array}{cc}
\left( \stackrel{,}{b}+\sqrt{1-\stackrel{,}{a}^2}\right) 1_4 & \left( -i%
\stackrel{,}{c}+\stackrel{,}{g}\right) \gamma _5 \\ 
\left( i\stackrel{,}{c}+\stackrel{,}{g}\right) \gamma _5 & \left( \sqrt{1-%
\stackrel{,}{a}^2}-\stackrel{,}{b}\right) 1_4
\end{array}
\right] \mbox{,} 
\]

\[
\stackrel{,}{\pi }_{*}=\frac 1{2\sqrt{1-\stackrel{,}{a}^2}}\left[ 
\begin{array}{cc}
\left( \sqrt{1-\stackrel{,}{a}^2}-\stackrel{,}{b}\right) 1_4 & \left( i%
\stackrel{,}{c}-\stackrel{,}{g}\right) \gamma _5 \\ 
\left( -\stackrel{,}{g}-i\stackrel{,}{c}\right) \gamma _5 & \left( \stackrel{%
,}{b}+\sqrt{1-\stackrel{,}{a}^2}\right) 1_4
\end{array}
\right] \mbox{.} 
\]

Let:

\[
\cos \stackrel{,}{\lambda }=\stackrel{,}{a}\mbox{ and }\sin \stackrel{,}{%
\lambda }=\sqrt{1-\stackrel{,}{a}^2} 
\]

and

\begin{equation}
\begin{array}{c}
\Psi `\rightarrow \Psi ^{\prime }=\left( \stackrel{,}{U}\Psi `\right) %
\mbox{,} \\ 
\underline{n}`\rightarrow \underline{n}^{\prime }=\left( \underline{n}`\cos 
\stackrel{,}{\lambda }+\underline{k}`\sin \stackrel{,}{\lambda }\left( 
\stackrel{,}{\pi }_{\circ }-\stackrel{,}{\pi }_{*}\right) \right) \mbox{,}
\\ 
\underline{k}`\rightarrow \underline{k}^{\prime }=\left( \underline{k}`\cos 
\stackrel{,}{\lambda }-\underline{n}`\sin \stackrel{,}{\lambda }\left( 
\stackrel{,}{\pi }_{\circ }-\stackrel{,}{\pi }_{*}\right) \right) \mbox{,}
\\ 
F_\mu \rightarrow F_\mu ^{\prime }=F_\mu \mbox{,} \\ 
B_\mu \rightarrow B_\mu ^{\prime }=B_\mu \mbox{,}
\end{array}
\label{tt2}
\end{equation}

and

\[
W_\mu \rightarrow W_\mu ^{\prime }\mbox{.} 
\]

In that case from (\ref{w}):

\[
W_\mu ^{\prime }=-\frac{2i}{g_2}\left( \partial _\mu \left( \stackrel{,}{U}%
U^{\left( -\right) }\right) \right) \left( \stackrel{,}{U}U^{\left( -\right)
}\right) ^{\dagger }; 
\]

Hence:

\[
W_\mu ^{\prime }=-\frac{2i}{g_2}\left( \partial _\mu \stackrel{,}{U}\right) 
\stackrel{,}{U}^{\dagger }-\frac{2i}{g_2}\stackrel{,}{U}\left( \partial _\mu
U^{\left( -\right) }\right) U^{\left( -\right) \dagger }\stackrel{,}{U}%
^{\dagger }\mbox{;} 
\]

i.e.:

\[
W_\mu ^{\prime }=\stackrel{,}{U}W_\mu \stackrel{,}{U}^{\dagger }-\frac{2i}{%
g_2}\left( \partial _\mu \stackrel{,}{U}\right) \stackrel{,}{U}^{\dagger }%
\mbox{.} 
\]

If

\[
F_{\mu ,\nu }=\left( \partial _\mu W_\nu -\partial _\nu W_\mu -i\frac{g_2}%
2\left( W_\mu W_\nu -W_\nu W_\mu \right) \right) 
\]

then

\[
F_{\mu ,\nu }^{\prime }=\partial _\mu W_\nu ^{\prime }-\partial _\nu W_\mu
^{\prime }-i\frac{g_2}2\left( W_\mu ^{\prime }W_\nu ^{\prime }-W_\nu
^{\prime }W_\mu ^{\prime }\right) =UF_{\mu ,\nu }U^{\dagger }\mbox{.} 
\]

Therefore $F_{\mu ,\nu }$ is invariant for the transformation (\ref{tt2}).

The Lagrangian for $F_{\mu ,\nu }$:

\[
\mathcal{L}_F=\left( -\frac 14\sum_{\mu ,\nu }F^{\mu ,\nu }F_{\mu ,\nu
}\right) \mbox{.} 
\]

Hence the Euler-Lagrange equations for $W_\mu $ are the following:

\[
\sum_\nu \partial ^\nu \left( \partial _\mu W_\nu -\partial _\nu W_\mu -i%
\frac{g_2}2\left[ W_\mu ,W_\nu \right] \right) =0\mbox{.} 
\]

For the components:

\[
\begin{array}{c}
\sum_\nu \partial ^\nu \partial _\nu W_{0,\mu }=g_2\sum_\nu \partial ^\nu
\left( W_{1,\mu }W_{2,\nu }-W_{2,\mu }W_{1,\nu }\right) +\partial _\mu
\sum_\nu \partial ^\nu W_{0,\nu }\mbox{,} \\ 
\sum_\nu \partial ^\nu \partial _\nu W_{1,\mu }=g_2\sum_\nu \partial ^\nu
\left( W_{0,\nu }W_{2,\mu }-W_{0,\mu }W_{2,\nu }\right) +\partial _\mu
\sum_\nu \partial ^\nu W_{1,\nu }\mbox{,} \\ 
\sum_\nu \partial ^\nu \partial _\nu W_{2,\mu }=g_2\sum_\nu \partial ^\nu
\left( W_{0,\mu }W_{1,\nu }-W_{0,\nu }W_{1,\mu }\right) +\partial _\mu
\sum_\nu \partial ^\nu W_{2,\nu }\mbox{.}
\end{array}
\]

Let:

\begin{equation}
\begin{array}{c}
\alpha _{0,\mu ,\nu }=\partial _\nu W_{0,\mu }-g_2\left( W_{1,\mu }W_{2,\nu
}-W_{2,\mu }W_{1,\nu }\right) \mbox{,} \\ 
\alpha _{1,\mu ,\nu }=\partial _\nu W_{1,\mu }-g_2\left( W_{0,\nu }W_{2,\mu
}-W_{0,\mu }W_{2,\nu }\right) \mbox{,} \\ 
\alpha _{2,\mu ,\nu }=\partial _\nu W_{2,\mu }-g_2\left( W_{0,\mu }W_{1,\nu
}-W_{0,\nu }W_{1,\mu }\right) \mbox{.}
\end{array}
\label{A1}
\end{equation}

Hence if $\sum_\nu \partial ^\nu W_\nu =0$ then

\[
\sum_\nu \partial ^\nu \alpha _{0,\mu ,\nu }=0,\sum_\nu \partial ^\nu \alpha
_{1,\mu ,\nu }=0,\sum_\nu \partial ^\nu \alpha _{2,\mu ,\nu }=0\mbox{.} 
\]

From (\ref{A1}):

\begin{equation}
\partial _\nu W_{0,\mu }=\left( g_2\left( W_{1,\mu }W_{2,\nu }-W_{2,\mu
}W_{1,\nu }\right) +\alpha _{0,\mu ,\nu }\right) \mbox{,}  \label{b1}
\end{equation}

\begin{equation}
\partial _\nu W_{1,\mu }=\left( g_2\left( W_{0,\nu }W_{2,\mu }-W_{0,\mu
}W_{2,\nu }\right) +\alpha _{1,\mu ,\nu }\right) \mbox{,}  \label{b2}
\end{equation}

\begin{equation}
\partial _\nu W_{2,\mu }=\left( g_2\left( W_{0,\mu }W_{1,\nu }-W_{0,\nu
}W_{1,\mu }\right) +\alpha _{2,\mu ,\nu }\right) \mbox{;}  \label{b3}
\end{equation}

from (\ref{b1}):

\begin{equation}
\begin{array}{c}
\partial _\nu \partial _\nu W_{0,\mu }=g_2\partial _\nu \left( W_{1,\mu
}W_{2,\nu }-W_{2,\mu }W_{1,\nu }\right) +\partial _\nu \alpha _{0,\mu ,\nu }=
\\ 
=g_2\left( \partial _\nu W_{1,\mu }W_{2,\nu }+W_{1,\mu }\partial _\nu
W_{2,\nu }-\partial _\nu W_{2,\mu }W_{1,\nu }-W_{2,\mu }\partial _\nu
W_{1,\nu }\right) +\partial _\nu \alpha _{0,\mu ,\nu };
\end{array}
\label{b4}
\end{equation}

hence from (\ref{b4}), (\ref{b2}) and (\ref{b3}):

\begin{eqnarray*}
\partial _\nu \partial _\nu W_{0,\mu } &=&g_2\left( 
\begin{array}{c}
\left( g_2\left( W_{0,\nu }W_{2,\mu }-W_{0,\mu }W_{2,\nu }\right) +\alpha
_{1,\mu ,\nu }\right) W_{2,\nu }- \\ 
-\left( g_2\left( W_{0,\mu }W_{1,\nu }-W_{0,\nu }W_{1,\mu }\right) +\alpha
_{2,\mu ,\nu }\right) W_{1,\nu }- \\ 
-W_{2,\mu }\partial _\nu W_{1,\nu }+W_{1,\mu }\partial _\nu W_{2,\nu }
\end{array}
\right) + \\
&&\ +\partial _\nu \alpha _{0,\mu ,\nu }\mbox{;}
\end{eqnarray*}

hence:

\[
\begin{array}{c}
\partial _\nu \partial _\nu W_{0,\mu }= \\ 
=g_2\left( 
\begin{array}{c}
g_2\left( -\left( W_{2,\nu }^2+W_{1,\nu }^2\right) W_{0,\mu }+\left(
W_{1,\mu }W_{1,\nu }+W_{2,\mu }W_{2,\nu }\right) W_{0,\nu }\right) + \\ 
+\alpha _{1,\mu ,\nu }W_{2,\nu }-\alpha _{2,\mu ,\nu }W_{1,\nu }+W_{1,\mu
}\partial _\nu W_{2,\nu }-W_{2,\mu }\partial _\nu W_{1,\nu }
\end{array}
\right) + \\ 
+\partial _\nu \alpha _{0,\mu ,\nu }\mbox{;}
\end{array}
\]

and

\[
\begin{array}{c}
\partial _\nu \partial _\nu W_{0,\mu }=-g_2^2\left( W_{2,\nu }^2+W_{1,\nu
}^2+W_{0,\nu }^2\right) W_{0,\mu }+ \\ 
+g_2^2\left( W_{0,\mu }W_{0,\nu }+W_{1,\mu }W_{1,\nu }+W_{2,\mu }W_{2,\nu
}\right) W_{0,\nu }+ \\ 
+g_2\left( \alpha _{1,\mu ,\nu }W_{2,\nu }-\alpha _{2,\mu ,\nu }W_{1,\nu
}+W_{1,\mu }\partial _\nu W_{2,\nu }-W_{2,\mu }\partial _\nu W_{1,\nu
}\right) +\partial _\nu \alpha _{0,\mu ,\nu }\mbox{;}
\end{array}
\]

if $\sum_\nu \partial _\nu W_\nu =0$ then:

\begin{equation}
\begin{array}{c}
\sum_\nu \partial _\nu \partial _\nu W_{0,\mu }=-g_2^2W_{0,\mu }\sum_\nu
W_\nu ^2+ \\ 
+\frac{g_2^2}2\sum_\nu \left( W_\mu W_\nu +W_\nu W_\mu \right) W_{0,\nu
}+g_2\sum_\nu \left( \alpha _{1,\mu ,\nu }W_{2,\nu }-\alpha _{2,\mu ,\nu
}W_{1,\nu }\right) \mbox{,}
\end{array}
\label{d1}
\end{equation}

\[
\begin{array}{c}
\sum_\nu \partial _\nu \partial _\nu W_{1,\mu }=-g_2^2W_{1,\mu }\sum_\nu
W_\nu ^2+ \\ 
+\frac{g_2^2}2\sum_\nu \left( W_\nu W_\mu +W_\mu W_\nu \right) W_{1,\nu
}+g_2\sum_\nu \left( W_{0,\nu }\alpha _{2,\mu ,\nu }-\alpha _{0,\mu ,\nu
}W_{2,\nu }\right)
\end{array}
\]

and

\[
\begin{array}{c}
\sum_\nu \partial _\nu \partial _\nu W_{2,\mu }=-g_2^2W_{2,\mu }\sum_\nu
W_\nu ^2+ \\ 
+\frac{g_2^2}2\sum_\nu \left( W_\nu W_\mu +W_\mu W_\nu \right) W_{2,\nu
}+g_2\sum_\nu \left( \alpha _{0,\mu ,\nu }W_{1,\nu }-W_{0,\nu }\alpha
_{1,\mu ,\nu }\right) \mbox{.}
\end{array}
\]

\[
\alpha _{\mu ,\nu }=\left[ 
\begin{array}{cc}
\alpha _{0,\mu ,\nu } & \alpha _{1,\mu ,\nu }-i\alpha _{2,\mu ,\nu } \\ 
\alpha _{1,\mu ,\nu }+i\alpha _{2,\mu ,\nu } & -\alpha _{0,\mu ,\nu }
\end{array}
\right] 
\]

then

\[
\begin{array}{c}
\sum_\nu \partial _\nu \partial _\nu W_\mu =-g_2^2W_\mu \sum_\nu W_\nu ^2+
\\ 
+\frac{g_2^2}2\sum_\nu \left( W_\nu W_\mu +W_\mu W_\nu \right) W_\nu -i\frac{%
g_2^2}2\sum_\nu \left( \alpha _{\mu ,\nu }W_\nu -W_\nu \alpha _{\mu ,\nu
}\right) \mbox{.}
\end{array}
\]

It is the motion equation for the field $W_\mu $ which has got a less than
unit 1 velocity. That is this field does not behave as a massless field.

Hence although $F_{\mu ,\nu }$ is a massless field but its components $W_\mu 
$ do not behave like a massless fields.

If

\[
\sum_\nu \left( W_\nu \frac{\partial W_\nu }{\partial W_\mu }+\frac{\partial
W_\nu }{\partial W_\mu }W_\nu \right) =0 
\]

then a real $\upsilon $ exists for which

\begin{equation}
\upsilon =\left( 2\sum_\nu W_\nu ^2\right) ^{\frac 12}  \label{mw}
\end{equation}

and

\[
\partial _{W_\mu }\upsilon =0 
\]

then the Lagrangian of $W_\mu $ is:

\[
\begin{array}{c}
\widehat{\mathcal{L}}=\sum_\nu \left( \partial _\nu W_\mu \right) \left(
\partial _\nu W_\mu \right) -g_2^2\frac{\upsilon ^2}2W_\mu ^2+ \\ 
+\frac{g_2^2}4\sum_\nu \left( W_\nu W_\mu +W_\mu W_\nu \right) ^2- \\ 
-i\frac{g_2^2}2\left( \left( \sum_\nu \left[ \alpha _{\mu ,\nu },W_\nu
\right] \right) W_\mu +W_\mu \left( \sum_\nu \left[ \alpha _{\mu ,\nu
},W_\nu \right] \right) \right) \mbox{.}
\end{array}
\]

It is a lagrangian of a field with mass

\[
M=g_2\frac \upsilon {\sqrt{2}} 
\]

and $M>0$.

\subparagraph{$A$ and $Z$ bosons}

Let $A_\mu $ and $Z_\mu $ are a fields for which \cite{KN}:

\begin{equation}
Z_\mu =\frac 1{\sqrt{g_1^2+g_2^2}}\left( g_2W_{0,\mu }-g_1B_\mu \right) %
\mbox{, }A_\mu =\frac 1{\sqrt{g_1^2+g_2^2}}\left( g_2B_\mu +g_1W_{0,\mu
}\right)  \label{c2}
\end{equation}

and

\begin{equation}
\sum_\nu \partial ^\nu \partial _\nu A_\mu =0\mbox{.}  \label{c3}
\end{equation}

Let denote:

\[
\frac{g_2^2}2\sum_\nu \left( W_\mu W_\nu +W_\nu W_\mu \right) W_{0,\nu
}+g_2\sum_\nu \left( \alpha _{1,\mu ,\nu }W_{2,\nu }-\alpha _{2,\mu ,\nu
}W_{1,\nu }\right) =\Lambda \mbox{.} 
\]

Hence from (\ref{d1}) and (\ref{mw}):

\begin{equation}
\sum_\nu \partial _\nu \partial _\nu W_{0,\mu }=-g_2^2\frac{\upsilon ^2}%
2W_{0,\mu }+\Lambda  \label{c4}
\end{equation}

From (\ref{c2}):

\begin{equation}
B_\mu =\frac 1{\sqrt{g_1^2+g_2^2}}\left( g_2A_\mu -g_1Z_\mu \right) \mbox{, }%
W_\mu ^0=\frac 1{\sqrt{g_1^2+g_2^2}}\left( g_1A_\mu +g_2Z_\mu \right) %
\mbox{.}  \label{c1}
\end{equation}

and

\[
\sum_\nu \partial ^\nu \partial _\nu A_\mu =\frac 1{\sqrt{g_1^2+g_2^2}%
}\left( g_2\sum_\nu \partial ^\nu \partial _\nu B_\mu +g_1\sum_\nu \partial
^\nu \partial _\nu W_{0,\mu }\right) \mbox{,} 
\]

from (\ref{c4}):

\[
\sum_\nu \partial ^\nu \partial _\nu A_\mu =\frac 1{\sqrt{g_1^2+g_2^2}%
}\left( 
\begin{array}{c}
g_2\left( \sum_\nu \partial ^\nu \partial _\nu B_\mu +g_1^2\frac{\upsilon ^2}%
2B_\mu -g_1^2\frac{\upsilon ^2}2B_\mu \right) + \\ 
+g_1\left( -g_2^2\frac{\upsilon ^2}2W_{0,\mu }+\Lambda \right)
\end{array}
\right) \mbox{,} 
\]

from (\ref{c1})

\begin{eqnarray*}
\sum_\nu \partial ^\nu \partial _\nu A_\mu &=&-\frac{\upsilon ^2}%
2g_1g_2\frac 1{g_1^2+g_2^2}\left( 2g_1g_2A_\mu +\left( g_2^2-g_1^2\right)
Z_\mu \right) + \\
&&\ \ +\frac 1{\sqrt{g_1^2+g_2^2}}\left( g_1\Lambda +g_2\left( \sum_\nu
\partial ^\nu \partial _\nu B_\mu +g_1^2\frac{\upsilon ^2}2B_\mu \right)
\right) \mbox{,}
\end{eqnarray*}

from (\ref{c3}):

\begin{equation}
\begin{array}{c}
A_\mu =-\left( g_2^2-g_1^2\right) \frac 1{2g_1g_2}Z_\mu + \\ 
+\frac 1{\upsilon ^2\left( g_1g_2\right) ^2}\sqrt{g_1^2+g_2^2}\left(
g_1\Lambda +g_2\left( \sum_\nu \partial ^\nu \partial _\nu B_\mu +g_1^2\frac{%
\upsilon ^2}2B_\mu \right) \right)
\end{array}
\label{q1}
\end{equation}

From (\ref{c2}):

\[
\sum_\nu \partial ^\nu \partial _\nu Z_\mu =\frac 1{\sqrt{g_1^2+g_2^2}%
}\left( g_2\sum_\nu \partial ^\nu \partial _\nu W_{0,\mu }-g_1\sum_\nu
\partial ^\nu \partial _\nu B_\mu \right) \mbox{,} 
\]

from (\ref{c4}):

\[
\sum_\nu \partial ^\nu \partial _\nu Z_\mu =\frac 1{\sqrt{g_1^2+g_2^2}%
}\left( 
\begin{array}{c}
g_2\left( -g_2^2\frac{\upsilon ^2}2W_{0,\mu }+\Lambda \right) - \\ 
-g_1\left( \sum_\nu \partial ^\nu \partial _\nu B_\mu +g_1^2\frac{\upsilon ^2%
}2B_\mu -g_1^2\frac{\upsilon ^2}2B_\mu \right)
\end{array}
\right) \mbox{,} 
\]

from (\ref{c1}):

\[
\sum_\nu \partial ^\nu \partial _\nu Z_\mu =-\frac{\upsilon ^2}2\frac
1{g_1^2+g_2^2}\left( g_1^4+g_2^4\right) Z_\mu -g_1g_2\frac{\upsilon ^2}%
2\frac 1{g_1^2+g_2^2}\left( g_2^2-g_1^2\right) A_\mu + 
\]

\[
+\frac 1{\sqrt{g_1^2+g_2^2}}\left( g_2\Lambda -g_1\left( \sum_\nu \partial
^\nu \partial _\nu B_\mu +g_1^2\frac{\upsilon ^2}2B_\mu \right) \right) 
\]

and from (\ref{q1}):

\[
\begin{array}{c}
\sum_\nu \partial ^\nu \partial _\nu Z_\mu =-\frac 12\frac{\upsilon ^2}%
2\left( g_1^2+g_2^2\right) Z_\mu + \\ 
+\frac 12\sqrt{g_1^2+g_2^2}\left( \frac 1{g_2}\Lambda -\frac 1{g_1}\left(
\sum_\nu \partial ^\nu \partial _\nu B_\mu +g_1^2\frac{\upsilon ^2}2B_\mu
\right) \right)
\end{array}
\]

That is $Z_\mu $ has got the mass:

\[
M_Z=\frac \upsilon 2\sqrt{g_1^2+g_2^2}\mbox{.} 
\]

\paragraph{Fragments}

Since

\[
\left( \pi _{\circ }+\pi _{*}\right) =1_8 
\]

then

\[
\begin{array}{c}
\sum_{\mu =0}^3\underline{\beta _\mu }i\left( \partial _\mu -i\frac
12g_2W_\mu \right) \left( \pi _{\circ }+\pi _{*}\right) \Psi `- \\ 
-h\left( 
\begin{array}{c}
\left( \underline{n}\cos \lambda \left( \pi _{\circ }+\pi _{*}\right) +%
\underline{k}\sin \lambda \left( \pi _{\circ }-\pi _{*}\right) \right)
\gamma + \\ 
+\left( \underline{k}\cos \lambda \left( \pi _{\circ }+\pi _{*}\right) -%
\underline{n}\sin \lambda \left( \pi _{\circ }-\pi _{*}\right) \right) \beta
\end{array}
\right) \Psi `+ \\ 
+\sum_{\mu =0}^3F_\mu \underline{\beta _\mu }\left( \pi _{\circ }+\pi
_{*}\right) \Psi `+ \\ 
+0.5g_1\underline{Y}\sum_{\mu =0}^3B_\mu \underline{\beta _\mu }\left( \pi
_{\circ }+\pi _{*}\right) \Psi `=0\mbox{.}
\end{array}
\]

Because

\[
\begin{array}{c}
\pi _{\circ }\beta =\beta \pi _{\circ }\mbox{, }\pi _{*}\beta =\beta \pi _{*}%
\mbox{,} \\ 
\pi _{\circ }\gamma =\gamma \pi _{\circ }\mbox{,}\pi _{*}\gamma =\gamma \pi
_{*}
\end{array}
\]

then

\[
\begin{array}{c}
\sum_{\mu =0}^3\underline{\beta _\mu }i\left( \partial _\mu -i\frac
12g_2W_\mu \right) \pi _{\circ }\Psi `+ \\ 
+\sum_{\mu =0}^3\underline{\beta _\mu }i\left( \partial _\mu -i\frac
12g_2W_\mu \right) \pi _{*}\Psi `- \\ 
-h\left( 
\begin{array}{c}
\underline{n}\cos \lambda \gamma \pi _{\circ }\Psi `+\underline{n}\cos
\lambda \gamma \pi _{*}\Psi `+ \\ 
+\underline{k}\sin \lambda \gamma \pi _{\circ }\Psi `-\underline{k}\sin
\lambda \gamma \pi _{*}\Psi `+ \\ 
+\underline{k}\cos \lambda \beta \pi _{\circ }\Psi `+\underline{k}\cos
\lambda \beta \pi _{*}\Psi `- \\ 
-\underline{n}\sin \lambda \beta \pi _{\circ }\Psi `+\underline{n}\sin
\lambda \beta \pi _{*}\Psi `
\end{array}
\right) + \\ 
+\sum_{\mu =0}^3F_\mu \underline{\beta _\mu }\pi _{\circ }\Psi `+\sum_{\mu
=0}^3F_\mu \underline{\beta _\mu }\pi _{*}\Psi `+ \\ 
+0.5g_1\underline{Y}\sum_{\mu =0}^3B_\mu \underline{\beta _\mu }\pi _{\circ
}\Psi `+ \\ 
+0.5g_1\underline{Y}\sum_{\mu =0}^3B_\mu \underline{\beta _\mu }\pi _{*}\Psi
`=0\mbox{.}
\end{array}
\]

Let

\[
\Psi _{\circ }`=\pi _{\circ }\Psi `\mbox{ and }\Psi _{*}`=\pi _{*}\Psi `%
\mbox{.} 
\]

In that case:

\begin{equation}
\begin{array}{c}
\sum_{\mu =0}^3\underline{\beta _\mu }i\left( \partial _\mu -i\frac
12g_2W_\mu \right) \Psi _{\circ }`- \\ 
-h\left( \left( \underline{n}\cos \lambda +\underline{k}\sin \lambda \right)
\gamma +\left( \underline{k}\cos \lambda -\underline{n}\sin \lambda \right)
\beta \right) \Psi _{\circ }`+ \\ 
+\sum_{\mu =0}^3F_\mu \underline{\beta _\mu }\Psi _{\circ }`+0.5g_1%
\underline{Y}\sum_{\mu =0}^3B_\mu \underline{\beta _\mu }\Psi _{\circ }`+ \\ 
\sum_{\mu =0}^3\underline{\beta _\mu }i\left( \partial _\mu -i\frac
12g_2W_\mu \right) \Psi _{*}`- \\ 
-h\left( \left( \underline{n}\cos \lambda -\underline{k}\sin \lambda \right)
\gamma +\left( \underline{k}\cos \lambda +\underline{n}\sin \lambda \right)
\beta \right) \Psi _{*}`+ \\ 
+\sum_{\mu =0}^3F_\mu \underline{\beta _\mu }\Psi _{*}`+0.5g_1\underline{Y}%
\sum_{\mu =0}^3B_\mu \underline{\beta _\mu }\Psi _{*}`=0\mbox{.}
\end{array}
\label{um5}
\end{equation}

Therefore for the (\ref{tt}) transformation from (\ref{fr1}, \ref{fr2}):

\[
\mathbf{\Psi }\left( x_5,x_4\right) =\mathbf{\Psi }_{\circ }\left(
x_5,x_4\right) +\mathbf{\Psi }_{*}\left( x_5,x_4\right) \rightarrow 
\]

\[
\rightarrow \mathbf{\Psi }`\left( x_5,x_4\right) =\mathbf{\Psi }_{\circ
}`\left( x_5,x_4\right) +\mathbf{\Psi }_{*}`\left( x_5,x_4\right) = 
\]

\[
=\frac 1{2\sqrt{1-a^2}}\cdot 
\]

\[
\cdot \left( 
\begin{array}{c}
\left( b+\sqrt{\left( 1-a^2\right) }\right) \left( \left( a+i\sqrt{1-a^2}%
\right) \phi _L\left( 0,0\right) +\phi _R\left( 0,0\right) \right) - \\ 
-\left( ic-g\right) \left( \left( a+i\sqrt{1-a^2}\right) \phi _L\left(
n,k\right) -\phi _R\left( n,k\right) \right) + \\ 
+\exp \left( -ih\left( \left( n\cos \lambda +k\sin \lambda \right)
x_5+\left( k\cos \lambda -n\sin \lambda \right) x_4\right) \right) \cdot \\ 
\cdot \left( 
\begin{array}{c}
\left( ic+g\right) \left( \left( a+i\sqrt{1-a^2}\right) \phi _L\left(
0,0\right) -\phi _R\left( 0,0\right) \right) + \\ 
+\left( b+\sqrt{\left( 1-a^2\right) }\right) \left( \left( a+i\sqrt{1-a^2}%
\right) \phi _L\left( n,k\right) +\phi _R\left( n,k\right) \right)
\end{array}
\right)
\end{array}
\right) + 
\]

\[
+\frac 1{2\sqrt{1-a^2}}\cdot 
\]

\[
\cdot \left( 
\begin{array}{c}
\left( \sqrt{\left( 1-a^2\right) }-b\right) \left( \left( a-i\sqrt{1-a^2}%
\right) \phi _L\left( 0,0\right) +\phi _R\left( 0,0\right) \right) + \\ 
+\left( ic-g\right) \left( \left( a-i\sqrt{1-a^2}\right) \phi _L\left(
n,k\right) -\phi _R\left( n,k\right) \right) + \\ 
+\exp \left( -ih\left( \left( n\cos \lambda -k\sin \lambda \right)
x_5+\left( k\cos \lambda +n\sin \lambda \right) x_4\right) \right) \cdot \\ 
\cdot \left( 
\begin{array}{c}
\left( g+ic\right) \left( -\left( a-i\sqrt{1-a^2}\right) \phi _L\left(
0,0\right) +\phi _R\left( 0,0\right) \right) + \\ 
+\left( b+\sqrt{\left( 1-a^2\right) }\right) \left( \left( a-i\sqrt{1-a^2}%
\right) \phi _L\left( n,k\right) +\phi _R\left( n,k\right) \right)
\end{array}
\right)
\end{array}
\right) \mbox{.} 
\]

That is:

\[
\mathbf{\Psi }\left( x_5,x_4\right) =\mathbf{\Psi }_{\circ }\left(
x_5,x_4\right) +\mathbf{\Psi }_{*}\left( x_5,x_4\right) \rightarrow 
\]

\[
\rightarrow \mathbf{\Psi }`\left( x_5,x_4\right) =\mathbf{\Psi }_{\circ
}`\left( x_5,x_4\right) +\mathbf{\Psi }_{*}`\left( x_5,x_4\right) = 
\]

\[
=\frac 1{2\sqrt{1-a^2}}\cdot 
\]

\[
\cdot \left( 
\begin{array}{c}
\left( b+\sqrt{\left( 1-a^2\right) }\right) \left( \left( a+i\sqrt{1-a^2}%
\right) \phi _L\left( 0,0\right) +\phi _R\left( 0,0\right) \right) - \\ 
-\left( ic-g\right) \left( \left( a+i\sqrt{1-a^2}\right) \phi _L\left(
n,k\right) -\phi _R\left( n,k\right) \right) + \\ 
+\exp \left( -ih\left( \left( na+k\sqrt{1-a^2}\right) x_5+\left( ka-n\sqrt{%
1-a^2}\right) x_4\right) \right) \cdot \\ 
\cdot \left( 
\begin{array}{c}
\left( ic+g\right) \left( \left( a+i\sqrt{1-a^2}\right) \phi _L\left(
0,0\right) -\phi _R\left( 0,0\right) \right) + \\ 
+\left( b+\sqrt{\left( 1-a^2\right) }\right) \left( \left( a+i\sqrt{1-a^2}%
\right) \varphi _L\left( n,k\right) +\phi _R\left( n,k\right) \right)
\end{array}
\right)
\end{array}
\right) + 
\]

\[
+\frac 1{2\sqrt{1-a^2}}\cdot 
\]

\[
\cdot \left( 
\begin{array}{c}
\left( \sqrt{\left( 1-a^2\right) }-b\right) \left( \left( a-i\sqrt{1-a^2}%
\right) \varphi _L\left( 0,0\right) +\phi _R\left( 0,0\right) \right) + \\ 
+\left( ic-g\right) \left( \left( a-i\sqrt{1-a^2}\right) \phi _L\left(
n,k\right) -\phi _R\left( n,k\right) \right) + \\ 
+\exp \left( -ih\left( \left( na-k\sqrt{1-a^2}\right) x_5+\left( ka+n\sqrt{%
1-a^2}\right) x_4\right) \right) \cdot \\ 
\cdot \left( 
\begin{array}{c}
\left( g+ic\right) \left( -\left( a-i\sqrt{1-a^2}\right) \phi _L\left(
0,0\right) +\phi _R\left( 0,0\right) \right) + \\ 
+\left( b+\sqrt{\left( 1-a^2\right) }\right) \left( \left( a-i\sqrt{1-a^2}%
\right) \phi _L\left( n,k\right) +\phi _R\left( n,k\right) \right)
\end{array}
\right)
\end{array}
\right) \mbox{.} 
\]

Let in some point $\left\langle t,x,y,z\right\rangle $ $\phi _L\left(
n,k\right) \neq \mathbf{0}$ or/and $\phi _R\left( n,k\right) \neq \mathbf{0}$

In that case (\ref{dl2}) in this point: $\phi _L\left( 0,0\right) =\mathbf{0}
$ and $\phi _R\left( 0,0\right) =\mathbf{0}$.

Hence:

\begin{equation}
\begin{array}{c}
\mathbf{\Psi }_{\circ }`\left( x_5,x_4\right) =\frac 1{2\sqrt{1-a^2}}\cdot
\\ 
\cdot \left( 
\begin{array}{c}
-\left( ic-g\right) \left( \left( a+i\sqrt{1-a^2}\right) \phi _L\left(
n,k\right) -\phi _R\left( n,k\right) \right) + \\ 
+\exp \left( -ih\left( \left( na+k\sqrt{1-a^2}\right) x_5+\left( ka-n\sqrt{%
1-a^2}\right) x_4\right) \right) \cdot \\ 
\cdot \left( b+\sqrt{\left( 1-a^2\right) }\right) \left( \left( a+i\sqrt{%
1-a^2}\right) \phi _L\left( n,k\right) +\phi _R\left( n,k\right) \right)
\end{array}
\right) \mbox {,}
\end{array}
\label{ff1}
\end{equation}

\begin{equation}
\begin{array}{c}
\mathbf{\Psi }_{*}`\left( x_5,x_4\right) =\frac 1{2\sqrt{1-a^2}}\cdot \\ 
\cdot \left( 
\begin{array}{c}
\left( ic-g\right) \left( \left( a-i\sqrt{1-a^2}\right) \phi _L\left(
n,k\right) -\phi _R\left( n,k\right) \right) + \\ 
+\exp \left( -ih\left( \left( na-k\sqrt{1-a^2}\right) x_5+\left( ka+n\sqrt{%
1-a^2}\right) x_4\right) \right) \cdot \\ 
\cdot \left( b+\sqrt{\left( 1-a^2\right) }\right) \left( \left( a-i\sqrt{%
1-a^2}\right) \phi _L\left( n,k\right) +\phi _R\left( n,k\right) \right)
\end{array}
\right)
\end{array}
\label{ff2}
\end{equation}

and

\begin{equation}
\begin{array}{c}
\mathbf{\Psi }`\left( x_5,x_4\right) = \\ 
=-i\left( ic-g\right) \phi _L\left( n,k\right) + \\ 
+\frac 1{2\sqrt{1-a^2}}\cdot \\ 
\cdot (\exp \left( -ih\left( \left( na+k\sqrt{1-a^2}\right) x_5+\left( ka-n%
\sqrt{1-a^2}\right) x_4\right) \right) \cdot \\ 
\cdot \left( b+\sqrt{\left( 1-a^2\right) }\right) \left( \left( a+i\sqrt{%
1-a^2}\right) \phi _L\left( n,k\right) +\phi _R\left( n,k\right) \right) +
\\ 
+\frac 1{2\sqrt{1-a^2}}\cdot \\ 
+\exp \left( -ih\left( \left( na-k\sqrt{1-a^2}\right) x_5+\left( ka+n\sqrt{%
1-a^2}\right) x_4\right) \right) \cdot \\ 
\cdot \left( b+\sqrt{\left( 1-a^2\right) }\right) \left( \left( a-i\sqrt{%
1-a^2}\right) \phi _L\left( n,k\right) +\phi _R\left( n,k\right) \right) %
\mbox {.}
\end{array}
\label{fff}
\end{equation}

\paragraph{Local probabilities}

Let:

\[
\begin{array}{c}
\Psi _{\circ }^{\dagger }\Psi _{\circ }=\rho _{\circ }\mbox {, }\Psi _{\circ
}`^{\dagger }\Psi _{\circ }`=\rho _{\circ }`\mbox {,} \\ 
\Psi _{\circ }^{\dagger }\underline{\beta _1}\Psi _{\circ }=j_{\circ x}%
\mbox
{, }\Psi _{\circ }`^{\dagger }\underline{\beta _1}\Psi _{\circ }`=j_{\circ
x}`\mbox {,} \\ 
\Psi _{\circ }^{\dagger }\underline{\beta _2}\Psi _{\circ }=j_{\circ y}%
\mbox
{, }\Psi _{\circ }`^{\dagger }\underline{\beta _2}\Psi _{\circ }`=j_{\circ
y}`\mbox {,} \\ 
\Psi _{\circ }^{\dagger }\underline{\beta _3}\Psi _{\circ }=j_{\circ z}%
\mbox
{, }\Psi _{\circ }`^{\dagger }\underline{\beta _3}\Psi _{\circ }`=j_{\circ
z}`\mbox {,}
\end{array}
\]

\[
\begin{array}{c}
\Psi _{*}^{\dagger }\Psi _{*}=\rho _{*}\mbox {, }\Psi _{*}`^{\dagger }\Psi
_{*}`=\rho _{*}`\mbox {,} \\ 
\Psi _{*}^{\dagger }\underline{\beta _1}\Psi _{*}=j_{*x}\mbox {, }\Psi
_{*}`^{\dagger }\underline{\beta _1}\Psi _{*}`=j_{*x}`\mbox {,} \\ 
\Psi _{*}^{\dagger }\underline{\beta _2}\Psi _{*}=j_{*y}\mbox {, }\Psi
_{*}`^{\dagger }\underline{\beta _2}\Psi _{*}`=j_{*y}`\mbox {,} \\ 
\Psi _{*}^{\dagger }\underline{\beta _3}\Psi _{*}=j_{*z}\mbox {, }\Psi
_{*}`^{\dagger }\underline{\beta _3}\Psi _{*}`=j_{*z}`\mbox {,}
\end{array}
\]

Because

\[
\begin{array}{c}
\left( \Psi _{\circ }^{\dagger }\Psi _{\circ }\right) ^2-\left( \left( \Psi
_{\circ }^{\dagger }\underline{\beta _1}\Psi _{\circ }\right) ^2+\left( \Psi
_{\circ }^{\dagger }\underline{\beta _2}\Psi _{\circ }\right) ^2+\left( \Psi
_{\circ }^{\dagger }\underline{\beta _3}\Psi _{\circ }\right) ^2\right) = \\ 
=\left( \Psi _{\circ }^{\dagger }\gamma \Psi _{\circ }\right) ^2+\left( \Psi
_{\circ }^{\dagger }\beta \Psi _{\circ }\right) ^2\mbox {,}
\end{array}
\]

\[
\begin{array}{c}
\left( \Psi _{*}^{\dagger }\Psi _{*}\right) ^2-\left( \left( \Psi
_{*}^{\dagger }\underline{\beta _1}\Psi _{*}\right) ^2+\left( \Psi
_{*}^{\dagger }\underline{\beta _2}\Psi _{*}\right) ^2+\left( \Psi
_{*}^{\dagger }\underline{\beta _3}\Psi _{*}\right) ^2\right) = \\ 
=\left( \Psi _{*}^{\dagger }\gamma \Psi _{*}\right) ^2+\left( \Psi
_{*}^{\dagger }\beta \Psi _{*}\right) ^2
\end{array}
\]

then the local densities are:

\[
\begin{array}{c}
\rho _{\circ o}^2=\rho _{\circ }^2-\left( j_{\circ x}^2+j_{\circ
y}^2+j_{\circ z}^2\right) =\left( \Psi _{\circ }^{\dagger }\gamma \Psi
_{\circ }\right) ^2+\left( \Psi _{\circ }^{\dagger }\beta \Psi _{\circ
}\right) ^2 \\ 
\rho _{\circ o}`^2=\rho _{\circ }`^2-\left( j_{\circ x}`^2+j_{\circ
y}`^2+j_{\circ z}`^2\right) = \\ 
=\left( \Psi _{\circ }`^{\dagger }\gamma \Psi _{\circ }`\right) ^2+\left(
\Psi _{\circ }`^{\dagger }\beta \Psi _{\circ }`\right) ^2
\end{array}
\]

\[
\begin{array}{c}
\rho _{*o}^2=\rho _{*}^2-\left( j_{*x}^2+j_{*y}^2+j_{*z}^2\right) =\left(
\Psi _{*}^{\dagger }\gamma \Psi _{*}\right) ^2+\left( \Psi _{*}^{\dagger
}\beta \Psi _{*}\right) ^2 \\ 
\rho _{*o}`^2=\rho _{*}`^2-\left( j_{*x}`^2+j_{*y}`^2+j_{*z}`^2\right) = \\ 
=\left( \Psi _{*}`^{\dagger }\gamma \Psi _{*}`\right) ^2+\left( \Psi
_{*}`^{\dagger }\beta \Psi _{*}`\right) ^2
\end{array}
\]

Let us design:

\[
\begin{array}{c}
\gamma _{\circ }=\pi _{\circ }^{\dagger }\gamma \pi _{\circ }\mbox{, }\gamma
_{*}=\pi _{*}^{\dagger }\gamma \pi _{*}\mbox{,} \\ 
\beta _{\circ }=\pi _{\circ }^{\dagger }\beta \pi _{\circ }\mbox{, }\beta
_{*}=\pi _{*}^{\dagger }\beta \pi _{*}\mbox{,}
\end{array}
\]

\[
\begin{array}{c}
\gamma _{\circ }`=U^{\left( -\right) \dagger }\gamma _{\circ }U^{\left(
-\right) }\mbox{, }\gamma _{*}`=U^{\left( -\right) \dagger }\gamma
_{*}U^{\left( -\right) }\mbox{,} \\ 
\beta _{\circ }`=U^{\left( -\right) \dagger }\beta _{\circ }U^{\left(
-\right) }\mbox{, }\beta _{*}`=U^{\left( -\right) \dagger }\beta
_{*}U^{\left( -\right) }\mbox{.}
\end{array}
\]

Since

\[
\begin{array}{c}
\gamma _{\circ }`=a\gamma _{\circ }+\sqrt{1-a^2}\beta _{\circ }\mbox{, }%
\beta _{\circ }`=a\beta _{\circ }-\sqrt{1-a^2}\gamma _{\circ }\mbox{,} \\ 
\gamma _{*}`=a\gamma _{*}-\sqrt{1-a^2}\beta _{*}\mbox{, }\beta _{*}`=a\beta
_{*}+\sqrt{1-a^2}\gamma _{*}
\end{array}
\]

then

\[
\rho _{\circ o}^2=\rho _{\circ o}`^2\mbox { and }\rho _{*o}^2=\rho _{*o}`^2%
\mbox {.} 
\]

From (\ref{uni}) since:

\[
\rho =\rho `\mbox {, }j_x=j_x`\mbox {, }j_y=j_y`\mbox {, }j_z=j_z` 
\]

then the local densities:

\[
\rho _o^2=\rho ^2-\left( j_x^2+j_y^2+j_z^2\right) =\rho _o`^2=\rho
`^2-\left( j_x`^2+j_y`^2+j_z`^2\right) \mbox {.} 
\]

Because

\[
\begin{array}{c}
\left( \Psi ^{\dagger }\Psi \right) ^2-\left( \left( \Psi ^{\dagger }%
\underline{\beta _1}\Psi \right) ^2+\left( \Psi ^{\dagger }\underline{\beta
_2}\Psi \right) ^2+\left( \Psi ^{\dagger }\underline{\beta _3}\Psi \right)
^2\right) = \\ 
=\left( \Psi ^{\dagger }\gamma \Psi \right) ^2+\left( \Psi ^{\dagger }\beta
\Psi \right) ^2
\end{array}
\]

and

\[
\begin{array}{c}
\left( \Psi `^{\dagger }\Psi `\right) ^2-\left( \left( \Psi `^{\dagger }%
\underline{\beta _1}\Psi `\right) ^2+\left( \Psi `^{\dagger }\underline{%
\beta _2}\Psi `\right) ^2+\left( \Psi `^{\dagger }\underline{\beta _3}\Psi
`\right) ^2\right) = \\ 
=\left( 
\begin{array}{c}
\sqrt{\left( \Psi _{\circ }`^{\dagger }\gamma \Psi _{\circ }`\right)
^2+\left( \Psi _{\circ }`^{\dagger }\beta \Psi _{\circ }`\right) ^2}+ \\ 
+\sqrt{\left( \Psi _{*}`^{\dagger }\gamma \Psi _{*}`\right) ^2+\left( \Psi
_{*}`^{\dagger }\beta \Psi _{*}`\right) ^2}
\end{array}
\right) ^2
\end{array}
\]

but

\[
\begin{array}{c}
\left( \Psi `^{\dagger }\Psi `\right) ^2-\left( 
\begin{array}{c}
\left( \Psi `^{\dagger }\underline{\beta _1}\Psi `\right) ^2+ \\ 
+\left( \Psi `^{\dagger }\underline{\beta _2}\Psi `\right) ^2+\left( \Psi
`^{\dagger }\underline{\beta _3}\Psi `\right) ^2
\end{array}
\right) \neq \\ 
\neq \left( \Psi `^{\dagger }\gamma \Psi `\right) ^2+\left( \Psi `^{\dagger
}\beta \Psi `\right) ^2
\end{array}
\]

then

\[
\begin{array}{c}
\rho _o=\sqrt{\left( \Psi ^{\dagger }\gamma \Psi \right) ^2+\left( \Psi
^{\dagger }\beta \Psi \right) ^2}= \\ 
=\sqrt{\left( \Psi _{\circ }{}^{\dagger }\gamma \Psi _{\circ }\right)
^2+\left( \Psi _{\circ }{}^{\dagger }\beta \Psi _{\circ }\right) ^2}+ \\ 
+\sqrt{\left( \Psi _{*}{}^{\dagger }\gamma \Psi _{*}\right) ^2+\left( \Psi
_{*}{}^{\dagger }\beta \Psi _{*}\right) ^2}= \\ 
=\rho _{\circ o}+\rho _{*o}
\end{array}
\]

and

\[
\begin{array}{c}
\rho _o`= \\ 
=\sqrt{\left( \Psi _{\circ }`^{\dagger }\gamma \Psi _{\circ }`\right)
^2+\left( \Psi _{\circ }`^{\dagger }\beta \Psi _{\circ }`\right) ^2}+ \\ 
+\sqrt{\left( \Psi _{*}`^{\dagger }\gamma \Psi _{*}`\right) ^2+\left( \Psi
_{*}`^{\dagger }\beta \Psi _{*}`\right) ^2}= \\ 
=\rho _{\circ o}`+\rho _{*o}`
\end{array}
\]

but

\[
\rho _o`^2\neq \left( \Psi `^{\dagger }\gamma \Psi `\right) ^2+\left( \Psi
`^{\dagger }\beta \Psi `\right) ^2\mbox {.} 
\]

Therefore $\rho _o$ is a local probability density of a sum of two mutually
exclusive events with a local densities $\rho _{\circ o}$ and $\rho _{*o}$.

Because:

\[
\begin{array}{c}
\Psi _{\circ }^{\dagger }\gamma \Psi _{\circ }=\frac 12\left( 1-\frac b{%
\sqrt{1-a^2}}\right) \Psi ^{\dagger }\gamma \Psi \mbox {,} \\ 
\Psi _{\circ }^{\dagger }\beta \Psi _{\circ }=\frac 12\left( 1-\frac b{\sqrt{%
1-a^2}}\right) \Psi ^{\dagger }\beta \Psi \mbox {,} \\ 
\Psi _{*}^{\dagger }\gamma \Psi _{*}=\frac 12\left( 1+\frac b{\sqrt{1-a^2}%
}\right) \Psi ^{\dagger }\gamma \Psi \mbox {,} \\ 
\Psi _{*}^{\dagger }\beta \Psi _{*}=\frac 12\left( 1+\frac b{\sqrt{1-a^2}%
}\right) \Psi ^{\dagger }\beta \Psi
\end{array}
\]

then $\rho _o$ and $\rho _o`$ do not depend from $U^{\left( -\right) }$.

For $U^{\left( +\right) }$ (\ref{upls}):

\[
\underline{\pi }_{\circ }=\frac 1{2\sqrt{1-u^2}}\left[ 
\begin{array}{cc}
\left( v+\sqrt{1-u^2}\right) 1_4 & \left( -s+ik\right) \gamma _5 \\ 
\left( -ik-s\right) \gamma _5 & \left( \sqrt{1-u^2}-v\right) 1_4
\end{array}
\right] \mbox{,} 
\]

\[
\underline{\pi }_{*}=\frac 1{2\sqrt{1-u^2}}\left[ 
\begin{array}{cc}
\left( \sqrt{1-u^2}-v\right) 1_4 & \left( s-ik\right) \gamma _5 \\ 
\left( ik+s\right) \gamma _5 & \left( v+\sqrt{1-u^2}\right) 1_4
\end{array}
\right] \mbox{.} 
\]

Hence:

\[
\begin{array}{c}
U^{\left( +\right) \dagger }\gamma U^{\left( +\right) }=u\gamma -\sqrt{1-u^2}%
\left( \underline{\pi }_{\circ }-\underline{\pi }_{*}\right) \beta \mbox{,}
\\ 
U^{\left( +\right) \dagger }\beta U^{\left( +\right) }=u\beta +\sqrt{1-u^2}%
\left( \underline{\pi }_{\circ }-\underline{\pi }_{*}\right) \gamma
\end{array}
\]

and all rest for $U^{\left( +\right) }$ like to $U^{\left( -\right) }$.

\end{document}